# Recent advances in high-dimensional mode-locked quantum frequency combs

Kai-Chi Chang[1,2,†,*], Xiang Cheng[1,2,†,**], Murat Can Sarihan[1,2], and Chee Wei Wong[1,2,***]

[1] Fang Lu Mesoscopic Optics and Quantum Electronics Laboratory, Department of Electrical and Computer Engineering, University of California, Los Angeles, CA 90095, USA

[2] Center for Quantum Science and Engineering, University of California, Los Angeles, CA 90095, USA

[†] These authors contributed equally

[*] Corresponding author emails: uclakcchang@ucla.edu (lead contact), chengxiang@ucla.edu, and cheewei.wong@ucla.edu

## Summary

High-dimensional entanglement in qudit states offers a promising pathway towards the realization of practical, large-scale quantum systems that are highly controllable. These systems can be leveraged for various applications, including advanced quantum information processing, secure communications, computation, and metrology. In this context, quantum frequency combs have a crucial role as they inherently support multiple modes in both temporal and frequency domains, while preserving a single spatial mode. The multiple temporal and frequency modes of quantum frequency combs facilitate the generation, characterization, and control of high-dimensional time-frequency entanglement in extensive quantum systems. In this review article, we provide an overview of recent technological advancements in high-dimensional energy-time entangled quantum frequency combs. We explore how these time-frequency qudits, achieved using scalable telecommunications-wavelength components, can empower the creation of large-scale quantum states. Advances in quantum frequency combs can unlock new capabilities and versatility for promising developments in quantum science and technology.

**Table of contents**

## I. Introduction

The qubit, a fundamental unit in quantum information processing, is a two-dimensional or two-state system. A qudit, representing a quantum system with $d$-dimensions, introduces a higher level of complexity due to the scaling of the Hilbert space dimensionality with $d^N$ (where $d$ and $N$ denote the number of dimensions and particles, respectively). This increased dimensionality aids the generation of large-scale quantum systems, serving as a foundation for essential tests of nature that aim to address fundamental loopholes [1-4] as well as practical quantum applications. The high-dimensional quantum states offer significant advantages, including a substantial boost in quantum information capacity, enhanced link speeds [5-10], improved resilience to noise in quantum communications [6, 10-14], advanced capabilities in quantum metrology [15, 16], and the facilitation of novel algorithms for quantum simulations and computations [17-20].

Various physical platforms offer unique capabilities for encoding and decoding large-scale quantum states, including atomic systems [21], superconducting devices [22], and photonic qudits [23]. In photonic qudit platforms, two fundamentally distinct methods are used to harness quantum information: continuous-variable and discrete-variable methods. The continuous-variable approach involves manipulating quantum information via the quantized electric field quadrature of squeezed states [24, 25], while the discrete-variable approach utilizes Fock states [26-28].

Continuous-variable systems encode quantum states in the continuous amplitude and phase of photons across different degrees of freedom (DoFs) and platforms [29]. The quadrature that experiences squeezing can be detected via homodyne detection [29, 30]. This approach enables the creation of large-scale photonic quantum systems, allowing for highly complex graph states and entangled networks with multiple modes in optical parametric oscillators (OPOs), which typically operate either above or below the threshold [30-39]. The main advantage of continuous-variable systems is the deterministic generation and room-temperature detection using homodyne photodiodes. However, OPOs are often bulky and require sophisticated stabilization [29-37]. To address this, continuous-variable squeezed quantum frequency combs have recently been demonstrated using on-chip platforms [40-44].

Discrete-variable systems, on the other hand, create and detect different degrees of freedom of single photons, allowing for the generation and precise control of individual modes in large-scale integrated quantum systems [45-48]. State-of-the-art superconducting nanowire single-photon detectors (SNSPDs) are used for coincidence counting measurements. The main advantage of discrete-variable systems is their ability to realize entangled states with near-unity fidelity and precise control over individual modes. However, current technology mainly relies on probabilistic entangled sources, posing scalability challenges as the number of qudits ($N$) increases. To overcome this, incorporating deterministic high-fidelity single-photon sources [49, 50], SNSPDs with near-unity detection efficiency [51-53], ultra-low dark counts [54], and minimal timing jitters [55] is essential. These advancements aim to achieve deterministic, unit-fidelity quantum information processing with single photons.

The initial creation and control of discrete-variable high-dimensional photonic systems have been successful in manipulating quantum information in spatial and orbital angular momentum modes [56-74]. However, the footprint and complexity of these approaches increase with the number of dimensions ($d$) or particles ($N$) of the qudit, primarily due to the spatial nature of the utilized DoFs, limiting the attainable complexity for practical systems.

In stark contrast, there has been a rapid investigation into the discretized temporal and spectral modes of mode-locked quantum frequency combs (QFC) [75, 76]. This trend marks a departure from previous approaches and has gained significant attention in recent studies [77-99]. It enables the coherent generation, characterization, and control of high-dimensional [79-84, 88, 94, 99], hyper-entangled [79, 86, 88, 99, 100], or multipartite states [86, 88] in a single spatial mode. The multimode temporal and spectral properties of mode-locked QFCs provide a unique opportunity for dense quantum information processing, making them well-suited for practical applications involving high-dimensional hyper-entangled quantum systems.

Here we review recent progress on the realization and applications of high-dimensional energy-time entangled mode-locked QFCs (see Fig. 1). Previous reviews have covered continuous-variable quantum information processing using squeezed states [24, 25, 29], and demonstrations utilizing other discrete-variable DoFs of entangled photons for complex quantum information processing and using integrated chips [102-106]. This review addresses the generation, characterization, and control of high-dimensional QFCs in the time and frequency domains. We focus on QFCs generated in nonlinear optical waveguides via post-filtering cavities and with cavity enhancement. Finally, we explore the potential of these $d$-level QFCs for future large-scale photonic quantum science and technology applications.

## II. High-dimensional time- and frequency-entanglement in quantum frequency combs

### A. Generation mechanism of a mode-locked quantum frequency comb

In this section, we focus on generating, characterizing, and control of discrete temporal and spectral modes within high-dimensional QFCs. We discuss methods for creating qudit states, techniques for their characterization, and strategies for their control. We particularly focus on the local and non-local measurements in QFCs with biphotons, which are of interest from both fundamental and application perspectives. Most configurations for creating entangled photon pairs in two or higher dimensions rely on the inherently probabilistic photon-pair generation processes, such as spontaneous parametric downconversion (SPDC) and spontaneous four-wave mixing. SPDC uses a $\chi^{(2)}$ nonlinearity to convert a higher-energy pump photon into two lower-energy photons (signal and idler) through three-wave mixing. Spontaneous four-wave mixing, however, is a third-order nonlinear effect where signal and idler photons are generated by the annihilation of two pump photons.

Mode-locked QFCs exhibit discrete spectral and temporal modes, and can be generated via SPDC photons with post-filtering, cavity enhancement, or spontaneous four-wave mixing in a microresonator [106]. In the time domain, the wavefunction of a QFC state is expressed as:

$$|\psi\rangle = \int d\tau e^{-\Delta\omega|\tau|} \sum_{m=-N_0}^{N_0} \text{sinc}(Am\Delta\Omega) \cos(m\Delta\Omega\tau) \hat{a}_H^\dagger(t) \hat{a}_V^\dagger(t+\tau)|0\rangle, \quad (1)$$

where the *sinc* function is the SPDC's phase-matching function with $A$ related to the phase-matching bandwidth; $\hat{a}_H^\dagger$ and $\hat{a}_V^\dagger$ are creation operators for horizontally and vertically polarized photons; $\Delta\Omega$ is the cavity FSR in rad s$^{-1}$; $\Omega$ is the detuning of the SPDC's biphotons from frequency degeneracy; $2N_0+1$ is the number of cavity lines passed by an overall bandwidth-limiting filter. The mode-locking mechanism, a well-established technique in ultrafast optics [107], is crucial for generating mode-locked QFCs. It is important to note the distinction in the term mode-locking between mode-locked laser pulses, classical frequency combs, and mode-locked QFCs [106]. In the classical domain, phase coherence among all mode-locked modes is essential to generate ultrafast pulses and classical frequency combs whereas, in the quantum regime, mode-locking refers primarily to phase stability between two photons in a QFC.

In 1999, Z. Y. Ou and Y. J. Lu observed the first quantum mode-locked biphoton state in a cavity-enhanced SPDC source using a Fabry-Perot cavity (Fig. 2a) [75]. The cavity was formed with the end facets of the nonlinear crystal as one reflecting surface. By filtering out non-degenerate pairs with a passive filter cavity, they measured the single-frequency-mode temporal distribution of biphotons and extracted a cavity bandwidth of 44 MHz (Fig. 2b). A few years later, the same group created another mode-locked two-photon state (Fig. 2c) using a fiber cavity to filter SPDC photons, discretizing their spectrum [76]. They observed the mode-locking process by measuring the first Hong-Ou-Mandel (HOM) revivals (Fig. 2d), as predicted by J. H. Shapiro [108], noting the HOM revival period was half the cavity round-trip time. In mode-locked QFC generation, biphotons with either a delay of one cavity round-trip time or simultaneous occurrence show coherence, both producing HOM interference [76]. In 2015, researchers observed high-visibility HOM revival interferences (central HOM dip visibility up to 96.5% after accidental subtractions) with 19 HOM dips, expanding usable temporal modes in quantum frequency combs (Fig. 2e) [79]. More recently, high-visibility HOM revival interferences were measured in a cavity-enhanced mode-locked biphoton state over a large path length difference [109]. Using a 100 GHz phase-matching bandwidth SPDC crystal, a 120.8 MHz FSR bow-tie cavity, and optical fiber with delays of multiple cavity round-trip times, they demonstrated HOM interferences up to the 84th dip, corresponding to a 105-m path length difference (Fig. 2f). The frequency spectrum of QFCs shows a multimode feature while conserving energy and momentum in the SPDC process, leading to discretized anti-correlation in the joint spectral intensity (JSI). Due to Fourier-transform duality, QFCs generate a strong multimode temporal correlation in the joint temporal intensity (JTI) when signal and idler photons are created simultaneously within the biphoton's coherence time.

## B. Characterizing the joint temporal intensity of spontaneous parametric downconversion

To fully utilize the multimode structure in QFCs, it is essential to measure and characterize the temporal and spectral properties of entangled photon pairs. We first discuss techniques for characterizing the joint temporal intensity (JTI) of SPDC two-photon states. Directly measuring JTI, which represents the probability of detecting biphotons at specific arrival times, is challenging with current SNSPDs due to the femtosecond timescales of SPDC photon correlations. Temporal cross-correlation measurements are feasible when the timing jitter of detectors (typically 20-200 ps) is smaller than the Fourier-transform of the cavity's free spectral range (FSR) for cavity-enhanced SPDC sources with few GHz FSR (Fig. 3a) [110] or the Fourier-transform bandwidth of SPDC biphotons in narrow phase-matching bandwidth sources [111]. Alternative methods to bypass this timescale limitation include sum-frequency generation (SFG) based ultrafast coincidence counting (Fig. 3b) [112-114] and temporal magnification with a time lens (Fig. 3c) [115]. Two-photon interferometry can reconstruct the temporal wavefunction of biphotons [116-118], and recently, conjugate Franson interferometry has been used to measure JTI and spectral phase of SPDC photons (Fig. 3d) [121, 122]. Generally, JTI measurements are limited by the temporal resolution of single-photon detectors. Besides JTI, direct measurement of the joint temporal amplitude (JTA) can provide phase information, but standard coincidence counting lacks tools for full spectral phase probing. JTA can be reconstructed via interferometric techniques [118, 120] or temporal amplitude-phase modulation with frequency-resolved detection [123].

## C. Characterizing the joint spectral intensity of spontaneous parametric downconversion

On the other hand, the joint spectral intensity (JSI), the conjugate of JTI, represents the frequency correlation of energy-matched photon pairs and can be measured via spectrally-resolved coincidence measurements with standard single-photon detectors [124]. We focus on common methods to characterize the JSI of SPDC biphoton pairs, including frequency-bin entanglement measurements (Fig. 3e) [75, 81, 84, 85, 94, 99, 100, 125], SFG-based ultrafast coincidence counting (Fig. 3f) [114, 126-128], dispersive fiber spectroscopy (Fig. 3g) [129-137], spectral magnification via time lens (Fig. 3h) [115, 138], and shifting biphoton frequencies by pump tuning or nonlinear crystal control [139]. Similar to JTI measurements, full quantum state reconstruction with coincidence counting methods is challenging, but direct measurement of joint spectral amplitude (JSA) has been demonstrated [118, 140-142], and new methods like modified Franson interferometry aim to retrieve full spectral amplitude and phase information [143]. Recently, a dispersion-engineered JSI SPDC source with a record-high 100 THz bandwidth has been demonstrated [144], and non-phase-matched SPDC sources offer a route to significantly increase the degree of frequency entanglement, surpassing standard phase-matched sources [145].

In addition to individual characterization of JTI and JSI, Franson and conjugate Franson interferometry can be used together to exploit temporal correlation and spectral anti-correlation [121, 148], enabling phase-sensitive quantum interference in Franson-type interferometers.

Franson interferometry uses unbalanced Mach-Zehnder interferometers (MZIs) with a path length difference ($\Delta T$) for the photon paths. To avoid local interference and accurately register indistinguishable single photons, $\Delta T$ must exceed the single-photon coherence time ($\sigma_{coh}$) and the timing jitter ($\delta T$) of the detectors. It incorporates a time delay and coincidence measurements to reveal the JSI of entangled photons [121]. Since Franson measurements are temporal, their visibility is sensitive to the temporal phase of biphotons [4, 79, 81, 99, 114]. Conjugate Franson interferometry uses two MZIs with equal path differences and a frequency shift ($\Delta\Omega$) to eliminate local interference, applying a frequency shift of $\pm\Delta\Omega$ and non-local dispersion cancellation $\pm\beta_2$ ($\beta_2\Delta\Omega > \delta T$) to measure frequency domain coincidences, revealing the JTI of biphotons [121, 122]. Combining both interferometry fully certifies energy-time entanglement by characterizing the biphoton wave function. While Franson interferometry measures JSI and temporal phase, conjugate Franson interferometry captures the spectral phase, providing a complete characterization of the quantum state. Dual-basis Franson interferometry can also be used for secure high-dimensional quantum key distribution (QKD) [121]. Further discussion on these techniques is in Section II.F.

## D. High-dimensional time-bin entanglement in quantum frequency combs

We explore high-dimensional entanglement in QFCs, focusing on the time and frequency domains. Besides local HOM revival interference measurements, Franson interferometry has shown the revival of quantum interference, with the fringe reappearance period corresponding to multiples of the QFC's inverse frequency spacing (Fig. 4a) [79, 81, 99, 149, 151]. These discrete temporal revivals, due to mode-locking, reveal high-dimensional energy-time entanglement. The temporal wavefunction of a QFC state has been described earlier in Eq. (1). Consequently, QFC scaling to $d$ dimensions can be achieved by increasing the FSR-to-bandwidth ratio of the cavity, allowing more time-bins with the same SPDC bandwidth. However, this leads to diminished Franson revival interference visibility due to the cavity's small finesse (Fig. 4a), reducing energy-time entanglement quality and resulting in a lower Schmidt number (Fig. 4b). This limits the scaling of entanglement of formation ($E_{of}$) and certifiable dimensionality in QFCs (Fig. 4c). The $E_{of}$, a certification of the dimensionality of the quantum frequency combs, is described by:

$$E_{of} \geq -\log_2\left(1 - \frac{B^2}{2}\right), \quad (2)$$

where

$$B = \frac{2}{\sqrt{|C|}}\left(\sum_{\substack{(j,k)\,\in\,C \\ j<k}} |\langle j,j|\rho|k,k\rangle| - \sqrt{\langle j,k|\rho|j,k\rangle\langle k,j|\rho|k,j\rangle}\right), \quad (3)$$

with $\rho$ being the time-binned state's density matrix, and $|j,k\rangle$ being the biphoton ket for the $j$th signal time-bin and the $k$th idler time-bin. Here $C$ is the set of time-bin indices used in the sum, with $|C|$ being that set's cardinality.

This lower bound remains useful even with access to only a submatrix of the density matrix. For a $d \times d$ submatrix, a maximally-entangled state has $B = \sqrt{2(d-1)/d}$, leading to $E_{of} = \log_2(d)$ ebits. The entanglement of formation quantifies entanglement and represents the minimum number of maximally entangled two-qubit states needed to create a density matrix $\rho$. In high-dimensional entanglement, this provides a quantitative lower bound, offering insights into the system's entanglement [150]. For qudit systems, $E_{of}$ must exceed 1 ebit to surpass the two-dimensional qubit limit. For measurements in Fig. 4a, $E_{of}$ of 1.89 ± 0.03 ebits is certified for a local dimension of 16 in a 45 GHz FSR QFC, and an $E_{of}$ of 1.40 ± 0.05 ebits is certified for the same local dimension in a 15 GHz FSR QFC [99]. Although ebits decline due to decreased Franson revival interference visibility, this can be improved with a cavity of higher finesse and reasonable $Q$-factor (typically $10^5$ to $10^7$ in telecom wavelengths) under the same FSR [90, 100, 151-153]. However, increasing cavity finesse improves Franson recurrence visibility but requires longer temporal delays for probing more time-bins. For even larger delays, the size of the fiber interferometer and its temperature stability become issues; a free-space bulky Franson interferometer [150, 154-156] could be used but needs active stabilization feedback for phase-sensitive non-local quantum interferences.

The scalability of high-dimensional time-bin entanglement is limited by detection capabilities, the number of interferometers, imbalances within interferometers, and the timing jitter of single-photon detectors. For instance, quantum state tomography for four-dimensional time-bin qudits has achieved an average state fidelity of 95% [157]. However, the number of required interferometers scales linearly with the number of time-bins or dimensionality $d$ [158]. Certifying high-dimensional or multipartite entanglement with a limited number of measurements is challenging, as quantum state tomography becomes extremely time-consuming or unavailable for complex states [159]. A common method is to use entanglement of formation with a reasonable number of measurements, and this technique has been applied to high-dimensional energy-time and time-bin entanglement in QFCs [88, 99, 153].

Another method to certify high-dimensional time-bin entanglement is through mutually unbiased basis (MUB) measurements [160-163]. In a Hilbert space of dimensionality $d$, MUBs consist of orthonormal bases where the inner product between states from different bases equals $1/d$. Measurements in one MUB provide no information about a state prepared in another, due to quantum uncertainty [163]. For a $d$-dimensional subspace, there are $d+1$ sets of MUBs [160]. Finding the correct MUBs is crucial for prepare-and-measure QKD schemes like BB84 [162, 164]. There are three main approaches: inserting equal amplitude but opposite sign group-velocity dispersions on biphotons [165, 166], using multiple Franson interferometers [158, 167], or employing a cascade of electro-optic phase modulators [168]. Each method has its strengths and challenges, such as insertion loss, alignment, and scaling [167]. Entanglement-based QKD protocols, however, do not require MUB measurements and use quantum non-local interferometry for entanglement characterization [6, 121, 169]. Franson-type interferometry can test Bell's inequality and set upper limits on Eve's Holevo information leakage [5, 146, 147]. While past work used single-photon detectors with timing jitter in the range of tens to hundreds of ps, recent advances with SNSPDs achieving sub-3 ps resolution [55] are expected to enhance the exploration and scaling of high-dimensional entanglement for applications like quantum communication and computing.

## E. High-dimensional frequency-bin entanglement in quantum frequency combs

In QFCs, mode-locking occurs simultaneously in both the time and frequency domains. Notably, the resulting spectral correlation from these quantum combs is discrete rather than continuous (Fig. 4d). Complementary to Eq. (1), the spectral wavefunction of the QFC is written as:

$$|\psi\rangle = \sum_{m=-N_0}^{N_0} \int d\Omega\, f'(\Omega)f(\Omega - m\Delta\Omega)\hat{a}_H^\dagger\left(\frac{\omega_p}{2}+\Omega\right)\hat{a}_V^\dagger\left(\frac{\omega_p}{2}-\Omega\right)|0\rangle, \quad (4)$$

where $f(\Omega - m\Delta\Omega)$ is the single frequency-bin profile defined by the cavity's Lorentzian transmission lineshape with FWHM linewidth $2\Delta\omega$, as $f(\Omega) = 1/[(\Delta\omega)^2 + \Omega^2]$. Since 2014, high-dimensional frequency-bin entanglement in QFCs has gained significant attention due to previous limitations in mixing and manipulating multiple frequency-bins are lifted. With the introduction of frequency-bin control [171], Fourier-transform pulse shapers and electro-optic phase modulators have become key tools for complex quantum information processing and quantum computing in QFCs [82, 83, 85-88, 94]. Pioneered by A. M. Weiner [107, 172], pulse shapers act as user-programmable filters for spectral amplitude and phase, while phase modulators facilitate frequency-bin superposition measurements. Cascading these tools allows for unitary and universal gate operations on multiple frequency-bins [85, 87, 88]. Besides generation and control, certifying high-dimensional frequency-bin entanglement is crucial. Solely measuring JSI for frequency-correlation (Fig. 4d) is inadequate due to unknown spectral phase coherence. Certification can be achieved by measuring spectral phase coherence via quantum interference after mixing frequency-bins (Fig. 4e) [92], or using JSI with complementary JTI measurements (Fig. 4f) [173]. Additionally, HOM and Franson revival interference measurements, combined with JTI measurements, can further confirm high-dimensional entanglement [149]. Schmidt mode decompositions are also used to quantify dimensionality by extracting the Schmidt number $K$ from the JSA or JTA. The Schmidt number $K$ is defined as:

$$K = (\sum \lambda_n^2)^{-1}, \text{where } \sum \lambda_n = 1, \quad (5)$$

with $\{\lambda_n\}$ being the Schmidt mode eigenvalues. For frequency- or time-bins, Schmidt eigenvalues come from the frequency-binned JSA or time-binned JTA. Directly measuring JSA and JTA is challenging due to the need to reconstruct complete phase information. Instead, JSI and JTI measurements are used to estimate the Schmidt number, assuming a pure quantum state [81, 82, 99]. The Schmidt number $K$ measures the number of correlated modes in a quantum state (Fig. 4g) [174, 175]. The discrete multimode nature of mode-locked QFCs allows quantification of their Hilbert space dimensionality through Schmidt mode decompositions in frequency and time domains [81, 82, 99]. Although the Schmidt number $K$ reflects frequency-correlation and correlated modes, it requires knowledge of the spectral phase from JSA. The frequency domain Schmidt number $K$ can represent quantum dimensionality in QFCs with direct JTI measurements or frequency-bin mixing. High-dimensional correlated frequency-bins has been demonstrated in QFCs [82, 84, 85, 92, 99], with a measured Schmidt number $K$ of approximately 20 in a 50 GHz FSR QFC and about 50 in a 25 GHz FSR QFC [84, 85]. Recently, an $E_{of}$ of 2.198 ± 0.007 ebits is certified in a 40 GHz FSR QFC using high-dimensional frequency-bin encoding and Bayesian tomography [92].

The scalability of high-dimensional frequency-bin entanglement is influenced by several factors, including the number of elements, bandwidth for electro-optic phase modulators, cavity FSR, and insertion loss per element (typically a few dB in telecom wavelengths). As with high-dimensional energy-time and time-bin entanglement, the number of elements increases linearly with the number of frequency-bins or dimensionality ($d$) of the entangled system. Modulation efficiency tends to decrease as dimensionality increases [176]. While commercially available devices typically have a bandwidth limit of 50 GHz, cutting-edge electro-optic phase modulators now reach 100 GHz [177, 178], with further improvements expected. Alternatively, using smaller FSR QFCs also scales the dimensionality d of high-dimensional frequency-bin entanglement. For instance, several studies have demonstrated high-dimensional frequency-bin entanglement with cavities having FSRs smaller than 50 GHz [85, 92, 99]. On-chip integration techniques can help mitigate coupling loss in electro-optic phase modulators, pulse shapers, and frequency shifters [177, 179, 180].

## F. Franson and conjugate Franson interferometry

Franson and conjugate Franson interferometry are key methods in quantum information science for characterizing and verifying the joint high-dimensional time-frequency entanglement, essential for advanced quantum communication and computing. These techniques use quantum superposition and interference principles to explore entanglement across multiple time- and frequency-bins, deepening our understanding of these quantum systems [12, 79, 81, 99, 122, 149-151].

Franson interferometry [148], introduced by James Franson in 1989, involves a two-path interferometer where a photon pair, generated via SPDC, is split and sent through two unbalanced MZIs (Fig. 5a). The path length difference introduces a relative time delay, enabling the measurement of two-photon interference patterns from indistinguishable temporal paths. This setup measures energy-time entanglement and coherence between photons. Recent demonstrations of Franson revival interferences [79] utilize tunable delays across multiple time-bins at the integer of the cavity, with central Franson visibility of 97.8% after accidental subtractions. This experiment provides the high-dimensional energy-time entanglement with interference recurrence visibilities in a QFC (Fig. 5b-5d). Conjugate Franson interferometry extends this approach by incorporating dispersive elements that introduce frequency-dependent delays [121, 122] (Fig. 5e). This allows control of interference patterns in the time domain and examination of entanglement in conjugate time-frequency variables, verifying spectral phase coherence with the same JSI but different JTI (Fig. 5f-5g). The current experimental conjugate Franson interference visibility is 96 ± 1% without background subtraction [122]. Moreover, the theoretical conjugate Franson recurrence interference visibilities for entangled QFC state is given in [99], providing another method for certification of high-dimensional frequency-bin entanglement in QFCs. These methods are crucial for probing quantum mechanics and have practical implications for quantum technologies. They enhance the performance and security of quantum communication, allowing greater information transmission per photon and resilience against eavesdropping [121]. Additionally, they are vital for developing quantum computing architectures using time-frequency modes as qudits, advancing large-scale quantum information processing.

## G. High-dimensional time-frequency entanglement in quantum frequency combs

Hyperentanglement involves correlation of all accessible DoFs between two or more particles, including time, frequency, spin angular momentum (polarization), and orbital angular momentum [181]. For example, the *d*-level quantum states involving multiple particles are often created using Greenberger-Horne-Zeilinger (GHZ) states with orbital angular momentum. These are typically generated in bulk free-space setups [66, 182], but suffer from reduced coherence time and detection rates, bounding the realization to 27-dimensional spaces [182]. In contrast, multimode time- and frequency-bin entanglement allows the creation of hyperentangled states, such as a four-party high-dimensional GHZ state [88] or a three-level four-partite cluster state [86], using only QFCs without extra particles. High-dimensional entanglement in quantum frequency combs involves connecting time-bin and frequency-bin observables, two discrete forms of energy-time entanglement. These states can be created by exciting SPDC or spontaneous four-wave mixing mediums with multiple temporal pulses (Fig. 6a). The key requirement is that pulse separation must exceed the cavity photon lifetime to achieve a time-frequency product greater than the EPR limit ($\Delta\nu\Delta t \gg 1$, where $\Delta\nu$, $\Delta t$ represent the frequency and time spacing between different modes), allowing independent control of time- and frequency-bins (Fig. 6b) [148, 183]. Time-frequency entanglement requires a pulsed pumping scheme, as continuous-wave pumping aligns cavity round-trip time with frequency spacing, yielding $\Delta\nu\Delta t = 1$. Recent demonstrations using a controlled-NOT gate and entangled states show a four-party GHZ state with 32 dimensions per DoF (Fig. 6c and 6d) [88], corresponding to a Hilbert dimensionality of $32^4$, equivalent to 20 qubits. However, this GHZ state, while realized using QFCs, is not directly applicable for genuine multi-party GHZ applications, such as Mermin inequality violations [182, 184].

Hyperentanglement by combining temporal or spectral modes with other DoFs is helpful for implementing deterministic controlled-NOT gates [185-190], foundational for applications such as dense quantum coding [79, 99, 191], high-dimensional quantum networks [192, 193], efficient quantum memory storage [194], and superdense quantum teleportation [195]. Specifically, entanglement in QFCs has been demonstrated for time-frequency [86], energy-time-polarization [79], and frequency-polarization DoFs [100]. High-dimensional time-frequency entanglement in QFCs leverages its multiple temporal and spectral mode structure, compatible with fiber optical technology. This approach enhances noise tolerance and increases the effective quantum resource rate, enabling advanced and robust quantum computations.

## III. Doubly- and singly-configured high-dimensional quantum frequency combs

### A. Doubly- and singly-resonant cavity-enhanced quantum frequency comb

The mode-locked QFCs discussed above operate in a doubly-resonant (DR) configuration, where the signal and idler photon spectra are confined within the cavity modes simultaneously (Fig. 2a). Since the first cavity-enhanced SPDC source was demonstrated [75], many studies have focused on DR-OPO schemes operated far below threshold [196-205]. This approach enhances the biphoton flux by a factor of cavity finesse and the brightness by the square of the finesse [206]. The bandwidth of these sources typically ranges from a few MHz to 100 MHz [207], useful for integrating quantum memory with atomic ensembles whose bandwidth is 10 to 100 MHz [208]. However, this scheme faces stability issues due to maintaining resonance conditions for both signal and idler beams under varying parameters. An improved method is using a triply-resonant SPDC source, which reduces the FSR and bandwidth, simplifying cavity locking and enhancing stability [109, 173, 209, 210].

On the other hand, singly-resonant (SR) QFCs have been proposed [206] and demonstrated in cavity-enhanced SPDC sources [90, 100, 211, 212]. Due to the entanglement between the signal and idler photons, both will display comb-like spectra [90, 100, 211], even if only one photon is resonant with the cavity mode. SR cavity-enhanced SPDC is more stable and tunable since only one photon interacts with the cavity. However, the brightness of an SR-OPO is typically lower than a DR-OPO, as the cavity only enhances one type of photon and does not amplify the total photon flux due to losses within the cavity. SR schemes also avoid the cluster effect common in DR-OPO configurations [90, 100], which suppresses photon pair generation due to the frequency-dependency of FSR [213, 214]. A recent review provides mathematical formulas and detailed implementations of cavity-enhanced SPDC sources [207]. Although the bandwidth of QFCs in SPDC sources is suitable for atomic quantum memories, the typical FSR is not compatible with commercial optical filters and modulators, limiting single-mode control and applications in quantum communications and computings.

## B. Doubly- and singly-filtered quantum frequency comb

Since 2015, versatile QFCs that can be configured in both doubly- and singly-filtered approaches have been investigated [79, 99, 149, 151]. In this setup, switching between these configurations can be done within a single experiment, avoiding the need to rebuild and stabilize the system (Fig. 7a). This is feasible for QFCs based on post-filtering, as the cavity mode structure can be applied to both biphotons simultaneously or to only signal or idler photons. The JSI shaping of type-II SPDC photons can be achieved by inserting a cavity before splitting the photons, which is the doubly-filtered configuration (Fig. 2c) [79, 99]. Conversely, changing the location of the splitting component enables singly-filtered operation by affecting only the spectrum of signal or idler photons, preserving the mode-locked QFC signature due to intrinsic photon pair entanglement [149, 151]. The JSI for both configurations is then compared (Fig. 7b). For doubly-filtered QFCs, its unnormalized frequency domain wavefunction is:

$$\psi_{Doubly-filtered}(\Omega) = \sum_{m=-N}^{N} \frac{\text{sinc}(A\Omega)}{|\Delta\omega + i(\Omega - m\Delta\Omega)|^2} \quad (6)$$

For the singly-filtered configuration, its unnormalized frequency domain wavefunction is:

$$\psi_{Singly-filtered}(\Omega) = \sum_{m=-N}^{N} \frac{\text{sinc}(A\Omega)}{\Delta\omega + i(\Omega - m\Delta\Omega)} \quad (7)$$

Although the FSR of the comb structure is the same for both doubly- and singly-filtered schemes, their JSI differs in two ways. First, in the doubly-filtered case, where both the signal and idler photons are filtered, the frequency-domain biphoton wave function decays faster with increasing frequency offset compared to the singly-filtered configuration. Second, the power spectrum differs:

the singly-filtered QFC has a higher unnormalized power spectral intensity than the doubly-filtered QFC (Fig. 7b). The JSI fall-off can be improved by using SPDC with a larger phase-matching bandwidth or a cavity with a smaller FSR. Singly-filtered QFCs also have higher photon flux than doubly-filtered ones, with no cavity enhancement for either scheme. This difference in JSI results in different JTI for the two post-filtering schemes. For the doubly-filtered scheme, the unnormalized temporal domain wavefunction is:

$$\Psi_{Doubly-filtered}(|n|\Delta T) = \exp(-|n|\Delta\omega\Delta T) \sum_{m=-N}^{N} \text{sinc}(Am\Delta\Omega) \qquad (8)$$

whereas for the singly-filtered configuration, its unnormalized temporal domain wavefunction is:

$$\Psi_{Singly-filtered}(n\Delta T) = \exp(-n\Delta\omega\Delta T) \sum_{m=-N}^{N} \text{sinc}(Am\Delta\Omega) \qquad (9)$$

The QFC state's temporal behavior exhibits recurrences with a period inverse to the comb spacing. Doubly-filtered QFCs show double-sided signal-idler cross-correlation $g^{(2)}$ functions [99], while singly-filtered QFCs show single-sided $g^{(2)}$ functions [149, 151, 206]. For a 45 GHz FSR fiber cavity, efficient generation of both doubly- and singly-filtered QFCs can be achieved. These exhibit double- and single-sided temporal oscillations (Fig. 7c) with a detection timing jitter $\delta T$ of 5 ps, achievable with state-of-the-art SNSPDs [55]. Recent observations using Franson interferometry have revealed high-dimensional energy-time entanglement in both QFC types (Fig. 7d) [99, 149, 151]. In singly-filtered QFCs, where only the signal photons pass through the cavity, Franson recurrences show faster fall-off for positive temporal delays compared to doubly-filtered QFCs. Despite the single-sided $g^{(2)}$ function, the Franson revival interferences cover both positive and negative delays due to the overlap integral with the delay-shifted counterpart [149]. The visibility of Franson recurrences at positive time-bins decreases faster than at negative time-bins, due to the asymmetric temporal profile of the cross-correlation function [149, 151]. The relationship between Franson revival visibility and time-bin Schmidt number $K_T$ is not fully understood, though recent investigations suggest a link with the two-photon JSI and corresponding Schmidt number [219, 220]. Understanding this connection involves analyzing the visibility of the $n^{th}$ Franson interference recurrence. The visibility of the $n^{th}$ Franson-interference recurrence in the doubly-filtered scheme is:

$$V_{n\,(Doubly-filtered)} = \exp(-|n|\Delta\omega\Delta T)(1 + |n|\Delta\omega\Delta T) \qquad (10)$$

Subsequently, the visibility of the positive $n^{th}$ Franson-interference recurrence for singly-filtered configuration can be described as:

$$V_{n\,(Singly-filtered)} = \exp(-n\Delta\omega\Delta T) \qquad (11)$$

where $\Delta T = 2\pi/\Delta\Omega$ is the repetition rate of QFCs. Then, the time-bin Schmidt eigenvalues $\lambda_n$ can be extracted via the following expression for doubly-filtered QFCs:

$$\lambda_{n\,(Doubly-filtered)} = \frac{e^{-\pi|n|/F}}{\sum_{n=-M}^{M} e^{-\pi|n|/F}}, 0 \leq |n| \leq M \qquad (12)$$

In singly-filtered QFC, the time-bin Schmidt eigenvalues $\lambda_n$ is:

$$\lambda_{n\,(Singly-filtered)} = \frac{e^{-\pi n/F}}{\sum_{n=0}^{M} e^{-\pi n/F}}, \text{ for } 0 \leq n \leq M \qquad (13)$$

where $M+1$ is the number of time-bins from Franson revival interferences, and $F$ is the cavity finesse. In both schemes, the time-bin Schmidt number $K_T$ can be extracted using Eq. (3). The symmetric and asymmetric Franson recurrence interferences in doubly- and singly-filtered schemes arise from their unnormalized temporal wave functions. We summarize the experimental Franson revival interference visibilities and time-bin Schmidt eigenvalue $\lambda_n$ in both configurations, with measured $K_T$ values of 13.11 (13.19 theoretical) and 20.72 (21.04), respectively (Fig. 7e). The temporal wavefunctions of QFCs suggest that improving the Franson visibility decay and $K_T$ could be achieved with a cavity of larger finesse $F$ given a fixed FSR [151].

In Table 1, we summarize recent key experimental QFC demonstrations from various configurations. Both the doubly- and singly-filtered configurations in QFCs offer their own set of advantages and disadvantages. For instance, the singly-filtered configuration exhibits a higher photon flux compared to the doubly-filtered configuration. As a result, it is more beneficial to employ the singly-filtered QFC for quantum communication applications such as high-dimensional entanglement distribution and wavelength-multiplexed QKD [149]. Notably, singly-filtered QFC can also be utilized for applications such as multimode quantum memory storage [204, 221-223]. The singly-filtered configuration provides increased stability and greater tunability, making it suitable for a broader range of non-degenerate or degenerate heralding measurements. In contrast, doubly-filtered QFC overall has better two-photon interference visibility and Hilbert space dimensionality, and is more suitable for time-frequency high-dimensional entanglement state generations towards cluster state quantum computations [86, 88] and for Sagnac interferometry [224, 225] with frequency-polarization entanglement in higher-dimensional dense encoding [79, 99].

The doubly- and singly-filtered QFCs offer greater versatility compared to cavity-enhanced SPDC sources. These mode-locked QFCs, implemented with post-filtering cavities, utilize telecom fiber components and do not necessitate active stabilization systems or intricate cavity designs. This makes them highly attractive as they simplify the experimental setup and eliminate the need for complex stabilization mechanisms. In contrast, the design of a cavity can be complex for cavity-enhanced SPDC sources in terms of cavity length, finesse, group velocities, and polarizations of biphotons. Moreover, it is straightforward to switch between doubly- and singly-filtered configurations by simply arranging an optical cavity in different places of the setups without redesigning and rebuilding the experimental setups. In contrast, implementing a similar arrangement in cavity-enhanced SPDC sources poses challenges. This is because the SPDC crystal or waveguide exhibits varying refractive indices for different wavelengths and polarizations of the signal and idler photons. Consequently, the effective cavity lengths differ for the respective resonance conditions, making it challenging to maintain consistent resonance conditions across the entire spectrum.

Compared to cavity-enhanced SPDC sources, although the brightness and photon flux are not enhanced in post-filtered QFCs, they offer a much less demanding flexible configuration for obtaining high-quality biphoton quantum interference fringes [79, 99, 149, 151], which is fundamental to advanced quantum information processing tasks. Although there will be unavoidable filtering loss for the generation of high-dimensional QFCs, the robust and stable fiber setups without sophisticated stabilization systems provide an alternate route toward generating large-scale photonic quantum systems for applications such as high-dimensional quantum communications and computations. In addition, for passive filtering QFCs, the FSR, and the cavity round-trip time can be chosen by any user to match the bandwidth of commercial optical bandpass filters, electro-optics phase-modulators and the frequency mode spacing of dense wavelength-division multiplexers, enabling single-mode control in the frequency domain. Combining with state-of-the-art SNSPDs with ultra-low timing jitter [55] and mature telecom interferometry technology, the complete and independent time- and frequency-bin control in QFCs is promising for applications ranging from scaling high-dimensional time-frequency entanglement, testing fundamental of quantum non-local nature, to real-world multi-user quantum networks.

## IV. Quantum information processing with quantum frequency combs
## A. High-dimensional quantum communications

The intrinsic temporal and spectral multimode structure in mode-locked QFCs is ideally suited for encoding and distributing qubits and qudits with telecom fiber networks. Indeed, for time-bin DoF, the first entanglement distribution experiment is implemented more than two decades ago [146]. Since then, time-bin has become a standard discrete-variable to perform entanglement distribution [147, 192, 226-230], and the high-dimensional encoding for QKD [5-7, 9, 10, 231-233]. In addition, several works implemented the idea of non-local dispersion cancelation [234, 235], using energy-time entangled biphotons [165, 166, 236, 237], whose technology is mutually compatible to mature dense wavelength-division multiplexing based QKD [82, 238-242]. Recently, the advanced high-dimensional entanglement distribution and frequency-multiplexed large-alphabet QKD in a QFC have been implemented using a singly-filtered scheme (Fig. 5a) [149]. Efficient entanglement distribution using DR-OPO configuration has also been shown (Fig. 5b) [243]. In particular, high-dimensional time-frequency entanglement distribution with 5 frequency-binned (averaged Franson visibility of 96.70 ± 1.93%, after accidental substractions) and 16 time-binned non-local interferences (central time-bin has a Franson visibility of 98.85 ± 0.50%, after accidental substractions) of a singly-filtered QFC at a 10-km distance at near zero-dispersion wavelength has been demonstrated (Fig. 8a) [149]. Proof-of-principle frequency-multiplexed time-bin high-dimensional QKD has also been demonstrated using singly- and doubly-filtered QFCs, with a singly-filtered QFC presents a ≈7.5× improvement on the secret key rate compared to a doubly-filtered QFC under same experimental conditions (Fig. 8a) [149]. A DR cavity-enhanced QFC has also been transmitted in a 10-km telecom fiber and then wavelength-converted from 1514 nm to 606 nm [243], for potential integration with efficient visible quantum memory (Fig. 8b). In general, the time-bin based high-dimensional entanglement distribution and QKD are more advantageous than qubit-based QKD for two main reasons. First, the intrinsic dense temporal encoding provides a higher key rate per photon [5-7, 9, 14, 244]; second, they exhibit higher tolerance of noise [10-13]. Moreover, frequency-multiplexed-based QKD offers advantages over qubit-based QKD. In frequency-multiplexed QKD, the coincidence rates on the detector side scale linearly with the number of utilized correlated pairs of channels. The primary limitations in this approach are the bandwidth of the SPDC source, the FSR of the cavity, the channel spacing of wavelength-division multiplexers, and the insertion loss associated with these devices.

Furthermore, it is proposed that combining the simultaneous usage of Franson and conjugate Franson interferometry can enable tighter constraints on the time-frequency covariance matrix of high-dimensional QKD [122]. This unconditional security proof is inspired by the continuous-variable QKD security proofs against Gaussian collective attacks, which use quadrature-component covariance matrices to derive Holevo bounds. It is suitable for qudits due to their ideal infinite Hilbert space. We should point out that Eve's temporal measurement decreases photon coherence while measuring frequency information increases biphoton correlation time. Therefore, by monitoring the perturbations in Franson and conjugate Franson interference visibilities using dual Bell inequalities, the time-frequency covariance matrix of the energy-time entangled photon pair can be more accurately constrained. This dual-basis protocol has been proposed to significantly enhance secure key rates, achieving up to 700 bits/s over a transmission distance of 200-km in optical fibers [122]. Furthermore, increasing the temporal path-length imbalance and amount of frequency shifting ensures a stricter bound, at the expense of higher phase stability requirements for dual-basis interferometry. For comprehensive reviews of high-dimensional QKD, see [103, 170, 245-247]. Overall, both the time-bin and frequency-bin-based high-dimensional QKD are rapidly investigated and improved regarding the key figure of merits such as secure key rates versus transmitted distance [6, 7, 9], and quantum bit error rates [5, 6, 248].

Future quantum communication protocols can be integrated with quantum memory via entanglement swapping and quantum teleportation to realize fully functional quantum repeaters [208]. For instance, the efficient quantum storage of frequency-multiplexed heralded single-photons in an atomic frequency comb using a DR-OPO operated at a quantum regime has been demonstrated at 606 nm [204]. By matching atomic frequency comb and QFC in their frequency domain, they store 15 discrete frequency modes separated by 261 MHz and spanning across 4 GHz (Fig. 8c) [204]. Furthermore, the robustness of multimode operation in both the time and frequency domains within a single spatial mode of QFCs holds potential benefits for quantum satellite applications [249-252], towards realizing next-generation global-scale quantum networks.

## B. High-dimensional quantum computation

Besides quantum communication, the qudits also have the advantage in quantum simulation and computation because of a reduction in the number of photons required to span and represent the arbitrary unitary matrix in a $d$-dimensional system compared to the qubits scenario [85, 253-255]. For example, qudits can implement quantum algorithms using smaller systems and fewer multi-site entangling gates [256]. Recently there has been theoretical proof that universal quantum computation can be realized using only linear optics and SPDC photons in any $d+1$ dimensional qudit basis [257]. Several studies have successfully demonstrated proof-of-principle quantum computational processing tasks utilizing high-dimensional entanglement in QFCs, including cluster state generations [30, 33, 37, 86, 258], linear optical quantum computation with passive devices [259], Bayesian tomography for frequency-bins in QFC [92], one-way $d$-level quantum computation via three-level four-partite QFC [86], and two-qudit quantum gate operations with fidelities exceeding 90% in the computational basis [88]. In particular, a high-dimensional multipartite quantum state with optimum entanglement witness and highest persistency of entanglement [260, 261], can be generated by transforming a time-frequency hyper-entangled state with a controlled phase gate for quantum computation (Fig. 8d) [86]. The QFCs based-cluster state is particularly useful for universal one-way quantum computing because all the information can be written, processed, and read out from the cluster by using only biphoton measurements (Fig. 8e and 8f) [261, 262]. The computational flow typically goes one-way, and feed-forward can be implemented between adjacent single-photon detections. Furthermore, the number of time-frequency hyper-entangled particles, the number of single-photons, as well as the dimensionality of the superposition can be further increased for higher computational performances toward a quantum advantage in the computational tasks [22, 263]. In Fig. 8g, we provide a summary of recent works that use QFCs for quantum communication and computation tasks. As the methodologies and technologies for high-dimensional entanglement continue to improve, the intrinsic large Hilbert space in temporal and spectral domains of QFCs provide an increasingly elegant way for advanced high-dimensional quantum computing.

Quantum interconnects, which enable the transfer of fragile coherent states between two designated devices, play a crucial role in distributed quantum information processing. These interconnects facilitate the reliable and efficient transfer of qubits and qudits while maintaining essential quantum properties such as superposition and entanglement, fundamental to quantum communication and computing in diverse platforms [264]. There are several types of quantum interconnects, depending on the physical systems used for communications and computation. Examples involve optical quantum interconnects that utilize flying photons in free-space, fiber-based, and chip-scale platforms to transfer quantum information; microwave quantum interconnects with superconducting circuits at cryogenic temperatures; spin-based quantum interconnects that utilize the electrons and spins in solid-state systems; and phononic quantum interconnects via lattice and mechanical vibrations. Hybrid quantum interconnects that combine different quantum technologies are also examined, enabling the faithful transfer of fragile quantum states between different physical segments or DoFs of the overall system [264]. Illustrative examples of hybrid quantum interconnects include, for instance, communication channels [239, 240, 245, 252], memories [194, 202-205, 221, 223], transduction between microwave and optical photons [265-271], conversion between different encoding schemes or DoFs [272-275], sensors [276], simulators [18, 277], and computation [26, 60, 278] in hybridized larger systems.

Quantifying key parameters in quantum interconnects involves measuring or estimating specific metrics that assess the link performance, efficiency, and reliability. For quantum optical interconnects with recent state-of-the-art QFCs, the time-frequency Hilbert space dimensionality $d^N$ of QFC provides an important role in terms of scalability, providing dense information [171, 180]. Furthermore, the QFC multiphoton quantum states have shown advantage in quantum computational processing compared to multiphoton states based on polarization DoF which is only two dimensional subspace [86]. For instance, QFCs from SR-OPO configuration has shown intrinsic photon-pair rate of $4\times10^6$, with over 1,000 frequency-bin pairs [90], and a three-level four-partite cluster state from QFC has shown orders-of-magnitude improvement in the effective quantum resource rate compared to cluster state based on polarization DoF [86]. Another key parameter in quantum optical interconnect is the fidelity of the photonic quantum state, where QFCs have achieved over 97% quantum state fidelity in Bell-state measurements and in Baysien tomography [279-281], and two-qudit gate fidelities over 90% [88]. In secure-key generation for quantum photonic interconnects, a singly-filtered QFC has presented a total key rate of ≈ 4.7 kbits/s for 5 frequency-bin pairs [149], with improvements in the source brightness further enhancing the rates. For quantum memories, another indispensable component of quantum optical interconnects, temporal and spectral multimodes can be stored and retrieved [194, 202-205, 282-287]. For example, a QFC based on DR-OPO has been stored in a atomic frequency comb quantum memory, with 15 discrete frequency-bins in a storage bandwidth of 4 GHz [204], and potential higher number of storage modes can be achieved by having broader memory bandwidth, or the narrower FSR of QFCs.

## V. Summary

The energy-time entangled QFCs discussed in this review present a robust and adaptable platform for generating high-dimensional hyperentangled states in a scalable manner. The advancements in the physics and engineering of the JSI and JTI from SPDC entangled sources enhances the achievable temporal and spectral high-dimensional entanglement in mode-locked QFCs. These high-dimensional entangled states offer a unique framework for manipulating large-scale quantum states within a single spatial mode, leveraging readily available telecommunications components. In high-dimensional time-bin entanglement, the QFC dimensionality is often bounded by the detector jitter and efficiencies, but steady SNSPD developments towards near-unity detection efficiency and low timing jitters have continued to advance the technology envelope. Simultaneously, high-dimensional frequency-bin entanglement is usually bounded by the FSR of the cavity, the operation bandwidth, and the insertion loss of the electro-optic components – these can be scaled by employing smaller FSR cavities, higher bandwidth and lower driving voltage modulators, and integrating lower loss devices.

Highly efficient and deterministic single-photon sources or stochastic photon-pair sources have and will significantly enhance the effectiveness of quantum photonic systems. These sources increase the photon flux, enabling the high-rate generation of higher-dimensional hyperentangled quantum states. Progress in observing and certifying high-dimensional quantum states, leveraging their multimode nature, continue to offer valuable assistance and guidance to experimentalists in fully capitalizing the capabilities of QFCs. Furthermore, as quantum technologies evolve, we expect the complete and independent time-frequency control of high-dimensional QFCs will be even more accessible in the near future, towards the generation, characterization, and control of arbitrary high-dimensional complex quantum states. These *d*-dimensional breakthroughs provide new possibilities for high-dimensional encoding schemes, quantum interconnect and memory storage, hardware-efficient quantum computation, sensing, and new protocols in quantum communication networks.

## Data Availability

The data that support the findings of this study are available from the corresponding author upon reasonable request.

## Declaration of Interest

The authors declare no competing interests.

## Acknowledgments

The authors thank useful discussions with J. H. Shapiro and F. N. C. Wong. This work is supported by the Army Research Office (W911NF-21-2-0214) and the National Science Foundation (QuIC-TAQS 2137984 and QII-TAQS 1936375).

## Author Contributions

K.-C.C., X.C., and C.W.W. initiated the review proposal. K.-C.C. and X.C. made substantial contributions to discussions of the content. K.-C.C., X.C., and M.C.S. researched data for the article. K.-C.C., X.C., M.C.S., and C.W.W. wrote the article and reviewed and/or edited the manuscript before submission.

The authors thank useful discussions with J. H. Shapiro and F. N. C. Wong. This work is supported by the Army Research Office (W911NF-21-2-0214) and the National Science Foundation (QuIC-TAQS 2137984 and QII-TAQS 1936375).

## References

[1] T. Vértesi, S. Pironio, and N. Brunner, Closing the detection loophole in Bell experiments using qudits, *Phys. Rev. Lett.* **104**, 060401 (2010).
[2] L. K. Shalm, E. Meyer-Scott, B. G. Christensen, P. Bierhorst, M. A. Wayne, M. J. Stevens, T. Gerrits, S. Glancy, D. R. Hamel, M. S. Allman, and K. J. Coakley, Strong loophole-free test of local realism, *Phys. Rev. Lett.* **115**, 250402 (2015).
[3] M. Giustina, M. A. Versteegh, S. Wengerowsky, J. Handsteiner, A. Hochrainer, K. Phelan, F. Steinlechner, J. Kofler, J. Å. Larsson, C. Abellán, and W. Amaya, Significant-loophole-free test of Bell's theorem with entangled photons, *Phys. Rev. Lett.* **115**, 250401 (2015).
[4] F. Vedovato, C. Agnesi, M. Tomasin, M. Avesani, J. Å. Larsson, G. Vallone, and P. Villoresi, Postselection-loophole-free Bell violation with genuine time-bin entanglement, *Phys. Rev. Lett.* **121**, 190401 (2018).
[5] I. Ali-Khan, C. J. Broadbent, and J. C. Howell, Large-alphabet quantum key distribution using energy-time entangled bipartite states, *Phys. Rev. Lett.* **98**, 060503 (2007).
[6] T. Zhong, H. Zhou, R. D. Horansky, C. Lee, V. B. Verma, A. E. Lita, A. Restelli, J. C. Bienfang, R. P. Mirin, T. Gerrits, S. W. Nam, F. Marsili, M. D. Shaw, Z. Zhang, L. Wang, D. Englund, G. W. Wornell, J. H. Shapiro, and F. N. C. Wong, Photon-efficient quantum key distribution using time-energy entanglement with high-dimensional encoding, *New J. Phys.* **17**, 022002 (2015).
[7] N. T. Islam, C. C. W. Lim, C. Cahall, J. Kim, and D. J. Gauthier, Provably secure and high-rate quantum key distribution with time-bin qudits, *Sci. Adv.* **3**, e1701491 (2017).
[8] M. Epping, H. Kampermann, and D. Bruß, Multi-partite entanglement can speed up quantum key distribution in networks, *New J. Phys.* **19**, 093012 (2017).

[9] C. Lee, D. Bunandar, Z. Zhang, G. R. Steinbrecher, P. B. Dixon, F. N. C. Wong, J. H. Shapiro, S. A. Hamilton, and D. Englund, Large-alphabet encoding for higher-rate quantum key distribution, *Opt. Express* **27**, 17539 (2019).

[10] H. Bechmann-Pasquinucci, and A. Peres, Quantum cryptography with 3-state systems, *Phys. Rev. Lett.* **85**, 3313 (2000).

[11]. J. Cerf, M. Bourennane, A. Karlsson, and N. Gisin, Security of quantum key distribution using *d*-level systems, *Phys. Rev. Lett.* **88**, 127902 (2002).

[12] S. Ecker, F. Bouchard, L. Bulla, F. Brandt, O. Kohout, F. Steinlechner, R. Fickler, M. Malik, Y. Guryanova, R. Ursin, and M. Huber, Overcoming noise in entanglement distribution, *Phys. Rev. X* **9**, 041042 (2019).

[13] F. Zhu, M. Tyler, N. H. Valencia, M. Malik, and J. Leach, Is high-dimensional photonic entanglement robust to noise? *AVS Quan. Sci.* **3**, 011401 (2021).

[14] M. Doda, M. Huber, G. Murta, M. Pivoluska, M. Plesch, and C. Vlachou, Quantum key distribution overcoming extreme noise: simultaneous subspace coding using high-dimensional entanglement, *Phys. Rev. Appl.* **15**, 034003 (2021).

[15] B. Koczor, S. Endo, T. Jones, Y. Matsuzaki, and S. C. Benjamin, Variational-state quantum metrology, *New J. Phys.* **22**, 083038 (2020).

[16] M. Tsang, F. Albarelli, and A. Datta, Quantum semiparametric estimation, *Phys. Rev. X* **10**, 031023 (2020).

[17] B. P. Lanyon, M. Barbieri, M. P. Almeida, T. Jennewein, T. C. Ralph, K. J. Resch, G. J. Pryde, J. L. O'brien, A. Gilchrist, and A. G. White, Simplifying quantum logic using higher-dimensional Hilbert spaces, *Nat. Phys.* **5**, 134 (2009).

[18] I. M. Georgescu, S. Ashhab, and F. Nori, Quantum simulation, *Rev. Mod. Phys.* **86**, 153 (2014).

[19] D.-S. Wang, D. T. Stephen, and R. Raussendorf, Qudit quantum computation on matrix product states with global symmetry, *Phys. Rev. A* **95**, 032312 (2017).

[20] S. McArdle, S. Endo, A. Aspuru-Guzik, S. C. Benjamin, and X. Yuan, Quantum computational chemistry, *Rev. Mod. Phys.* **92**, 015003 (2020).

[21] J. Zhang, G. Pagano, P. W. Hess, A. Kyprianidis, P. Becker, H. Kaplan, A. V. Gorshkov, Z.-X. Gong and C. Monroe, Observation of a many-body dynamical phase transition with a 53-qubit quantum simulator, *Nature* **551**, 601 (2017).

[22] F. Arute, K. Arya, R. Babbush, D. Bacon, J. C. Bardin, R. Barends, R. Biswas, S. Boixo, F. G. S. L. Brandao, D. A. Buell, B. Burkett, Y. Chen, Z. Chen, B. Chiaro, R. Collins, W. Courtney, A. Dunsworth, E. Farhi, B. Foxen, A. Fowler, C. Gidney, M. Giustina, R. Graff, K. Guerin, S. Habegger, M. P. Harrigan, M. J. Hartmann, A. Ho, M. Hoffmann, T. Huang, T. S. Humble, S. V. Isakov, E. Jeffrey, Z. Jiang, D. Kafri, K. Kechedzhi, J. Kelly, P. V. Klimov, S. Knysh, A. Korotkov, F. Kostritsa, D. Landhuis, M. Lindmark, E. Lucero, D. Lyakh, S. Mandrà, J. R. McClean, M. McEwen, A. Megrant, X. Mi, K. Michielsen, M. Mohseni, J. Mutus, O. Naaman, M. Neeley, C. Neill, M. Y. Niu, E. Ostby, A. Petukhov, J. C. Platt, C. Quintana, E. G. Rieffel, P. Roushan, N. C. Rubin, D. Sank, K. J. Satzinger, V. Smelyanskiy, K. J. Sung, M. D. Trevithick, A. Vainsencher, B. Villalonga, T. White, Z. J. Yao, P. Yeh, A. Zalcman, H. Neven and J. M. Martinis, Quantum supremacy using a programmable superconducting processor, *Nature* **574**, 505 (2019).

[23] J. M. Arrazola, V. Bergholm, K. Brádler, T. R. Bromley, M. J. Collins, I. Dhand, A. Fumagalli, T. Gerrits, A. Goussev, L. G. Helt, J. Hundal, T. Isacsson, R. B. Israel, J. Izaac, S. Jahangiri, R. Janik, N. Killoran, S. P. Kumar, J. Lavoie, A. E. Lita, D. H. Mahler, M. Menotti, B. Morrison, S. W. Nam, L. Neuhaus, H. Y. Qi, N. Quesada, A. Repingon, K. K. Sabapathy, M. Schuld, D. Su, J.

Swinarton, A. Száva, K. Tan, P. Tan, V. D. Vaidya, Z. Vernon, Z. Zabaneh and Y. Zhang, Quantum circuits with many photons on a programmable nanophotonic chip, *Nature* **591**, 54 (2021).
[24] S. L. Braunstein, and P. Van Loock, Quantum information with continuous variables, *Rev. Mod. Phys.* **77**, 513 (2005).
[25] C. Weedbrook, S. Pirandola, R. García-Patrón, N. J. Cerf, T. C. Ralph, J. H. Shapiro, and S. Lloyd, Gaussian quantum information, *Rev. Mod. Phys.* **84**, 621 (2012).
[26] P. Kok, W. J. Munro, K. Nemoto, T. C. Ralph, J. P. Dowling, and G. J. Milburn, Linear optical quantum computing with photonic qubits, *Rev. Mod. Phys.* **79**, 135 (2007).
[27] R. Horodecki, P. Horodecki, M. Horodecki, and K. Horodecki, Quantum entanglement, *Rev. Mod. Phys.* **81**, 865 (2009).
[28] J. W. Pan, Z. B. Chen, C. Y. Lu, H. Weinfurter, A. Zeilinger, and M. Żukowski, Multiphoton entanglement and interferometry, *Rev. Mod. Phys.* **84**, 777 (2012).
[29] K. Zhang, S. Liu, Y. Chen, X. Wang, and J. Jing, Optical quantum states based on hot atomic ensembles and their applications, *Photon. Insights* **1**, R06 (2022).
[30] J. I. Yoshikawa, S. Yokoyama, T. Kaji, C. Sornphiphatphong, Y. Shiozawa, K. Makino, and A. Furusawa, Invited article: Generation of one-million-mode continuous-variable cluster state by unlimited time-domain multiplexing, *APL Photon.* **1**. 060801 (2016).
[31] O. Pinel, P. Jian, R. M. De Araujo, J. Feng, B. Chalopin, C. Fabre, and N. Treps, Generation and characterization of multimode quantum frequency combs, *Phys. Rev. Lett.* **108**, 083601 (2012).
[32] S. Yokoyama, R. Ukai, S. C. Armstrong, C. Sornphiphatphong, T. Kaji, S. Suzuki, J.-I Yoshikawa, H. Yonezawa, N. C. Menicucci and A. Furusawa, Ultra-large-scale continuous-variable cluster states multiplexed in the time domain, *Nat. Photon.* **7**, 982 (2013).
[33] M. Chen, N. C. Menicucci, and O. Pfister, Experimental realization of multipartite entanglement of 60 modes of a quantum optical frequency comb, *Phys. Rev. Lett.* **112**, 120505 (2014).
[34] J. Roslund, R. M. de Araújo, S. Jiang, C. Fabre, and N. Treps, Wavelength-multiplexed quantum networks with ultrafast frequency combs, *Nat. Photon.* **8**, 109 (2014).
[35] S. Gerke, J. Sperling, W. Vogel, Y. Cai, J. Roslund, N. Treps, and C. Fabre, Full multipartite entanglement of frequency-comb Gaussian states, *Phys. Rev. Lett.* **114**, 050501 (2015).
[36] Y. Cai, J. Roslund, G. Ferrini, F. Arzani, X. Xu, C. Fabre and N. Treps, Multimode entanglement in reconfigurable graph states using optical frequency combs, *Nat. Commun.* **8**, 15645 (2017).
[37] X. Zhu, C.-H. Chang, C. González-Arciniegas, A. Pe'er, J. Higgins, and O. Pfister, Hypercubic cluster states in the phase-modulated quantum optical frequency comb, *Optica* **8**, 281 (2021).
[38] L. S. Madsen, F. Laudenbach, M. F. Askarani, F. Rortais, T. Vincent, J. F. F. Bulmer, F. M. Miatto, L. Neuhaus, L. G. Helt, M. J. Collins, A. E. Lita, T. Gerrits, S. W. Nam, V. D. Vaidya, M. Menotti, I. Dhand, Z. Vernon, N. Quesada and J. Lavoie, Quantum computational advantage with a programmable photonic processor, *Nature* **606**, 75 (2022).
[39] M. V. Larsen, X. Guo, C. R. Breum, J. S. Neergaard-Nielsen, and U. L. Andersen, Deterministic generation of a two-dimensional cluster state, *Science* **366**, 369 (2019).
[40] Y. Zhao, Y. Okawachi, J. K. Jang, X. Ji, M. Lipson, and A. L. Gaeta, Near-degenerate quadrature-squeezed vacuum generation on a silicon-nitride chip, *Phys. Rev. Lett.* **124**, 193601 (2020).
[41] V. D. Vaidya, B. Morrison, L. G. Helt, R. Shahrokshahi, D. H. Mahler, M. J. Collins, K. Tan, J. Lavoie, A. Repingon, M. Menotti, and N. Quesada, Broadband quadrature-squeezed vacuum

and nonclassical photon number correlations from a nanophotonic device, *Sci. Adv.* **6**, eaba9186 (2020).
[42] Z. Yang, M. Jahanbozorgi, D. Jeong, S. Sun, O. Pfister, H. Lee, and X. Yi, A squeezed quantum microcomb on a chip, *Nat. Commun.* **12**, 1 (2021).
[43] M. Jahanbozorgi, Z. Yang, S. Sun, H. Chen, R. Liu, B. Wang, and X. Yi, Generation of squeezed quantum microcombs with silicon nitride integrated photonic circuits, *Optica* **10**, 1100 (2023).
[44] M. A. Guidry, D. M. Lukin, K. Y. Yang, and J. Vučković, Multimode squeezing in soliton crystal microcombs, *Optica* **10**, 694 (2023).
[45] J. Wang, S. Paesani, Y. Ding, R. Santagati, P. Skrzypczyk, A. Salavrakos, J. Tura, R. Augusiak, L. Mancinska, D. Bacco, D. Bonneau, J. W. Silverstone, Q. Gong, A. Acin, K. Rottwitt, L. K. Oxenlowe, J. L. O'Brien, A. Laing, and M. G. Thompson, Multidimensional quantum entanglement with large-scale integrated optics, *Science* **360**, 285 (2018).
[46] X. Qiang, X. Zhou, J. Wang, C. M. Wilkes, T. Loke, S. O'Gara, L. Kling, G. D. Marshall, R. Santagati, T. C. Ralph, J. B. Wang, J. L. O'Brien, M. G. Thompson, and J. C. F. Matthews, Large-scale silicon quantum photonics implementing arbitrary two-qubit processing, *Nat. Photon.* **12**, 534 (2018).
[47] J. C. Adcock, C. Vigliar, R. Santagati, J. W. Silverstone, and M. G. Thompson, Programmable four-photon graph states on a silicon chip, *Nat. Commun.* **10**, 3528 (2019).
[48] D. Llewellyn, Y. Ding, I. I. Faruque, S. Paesani, D. Bacco, R. Santagati, Y.-J. Qian, Y. Li, Y.-F. Xiao, and M. Huber, M. Malik, G. F. Sinclair, X. Zhou, K. Rottwitt, J. L. O'Brien, J. G. Rarity, Q. Gong, L. K. Oxenlowe, J. Wang and M. G. Thompson, Chip-to-chip quantum teleportation and multi-photon entanglement in silicon, *Nat. Phys.* **16**, 148 (2019).
[49] H. Wang, Y.-M. He, T.-H. Chung, H. Hu, Y. Yu, S. Chen, X. Ding, M.-C. Chen, J. Qin, X. Yang, R.-Z. Liu, Z.-C. Duan, J.-P. Li, S. Gerhardt, K. Winkler, J. Jurkat, L.-J. Wang, N. Gregersen, Y.-H. Huo, Q. Dai, S. Yu, S. Höfling, C.-Y. Lu and J.-W. Pan, Towards optimal single-photon sources from polarized microcavities, *Nat. Photon.* **13**, 770 (2019).
[50] R. Uppu, H. T. Eriksen, H. Thyrrestrup, A. D. Uğurlu, Y. Wang, S. Scholz, A. D. Wieck, A. Ludwig, M. C. Löbl, R. J. Warburton, P. Lodahl and L. Midolo, On-chip deterministic operation of quantum dots in dual-mode waveguides for a plug-and-play single-photon source, *Nat. Commun.* **11**, 3782 (2020).
[51] F. Marsili, V. B. Verma, J. A. Stern, S. Harrington, A. E. Lita, T. Gerrits, I. Vayshenker, B. Baek, M. D. Shaw, R. P. Mirin and S. W. Nam, Detecting single infrared photons with 93% system efficiency, *Nat. Photon.* **7**, 210 (2013).
[52] D. V. Reddy, R. R. Nerem, S. W. Nam, R. P. Mirin, and V. B. Verma, Superconducting nanowire single-photon detectors with 98% system detection efficiency at 1550 nm, *Optica* **7**, 1649 (2020).
[53] J. Chang, J. W. N. Los, J. O. Tenorio-Pearl, N. Noordzij, R. Gourgues, A. Guardiani, J. R. Zichi, S. F. Pereira, H. P. Urbach, V. Zwiller, S. N. Dorenbos, and I. Esmaeil Zadeh, Detecting telecom single photons with 99.5-2.07+0.5% system detection efficiency and high time resolution, *APL Photon.* **6**, 036114 (2021).
[54] Y. Hochberg, I. Charaev, S. W. Nam, V. Verma, M. Colangelo, and K. K., Berggren, Detecting sub-GeV dark matter with superconducting nanowires, *Phys. Rev. Lett.* **123**, 151802 (2019).
[55] B. Korzh, Q.-Y. Zhao, J. P. Allmaras, S. Frasca, T. M. Autry, E. A. Bersin, A. D. Beyer, R. M. Briggs, B. Bumble, M. Colangelo, G. M. Crouch, A. E. Dane, T. Gerrits, A. E. Lita, F. Marsili,

G. Moody, C. Peña, E. Ramirez, J. D. Rezac, N. Sinclair, M. J. Stevens, A. E. Velasco, V. B. Verma, E. E. Wollman, S. Xie, D. Zhu, P. D. Hale, M. Spiropulu, K. L. Silverman, R. P. Mirin, S. W. Nam, A. G. Kozorezov, M. D. Shaw, and K. K. Berggren, Demonstration of sub-3 ps temporal resolution with a superconducting nanowire single-photon detector, *Nat. Photon.* **14**, 250 (2020).

[56] D. Bouwmeester, J. W. Pan, M. Daniell, H. Weinfurter, and A. Zeilinger, Observation of three-photon Greenberger-Horne-Zeilinger entanglement, *Phys. Rev. Lett.* **82**, 1345 (1999).

[57] A. Mair, A. Vaziri, G. Weihs, and A. Zeilinger, Entanglement of the orbital angular momentum states of photons, *Nature* **412**, 313 (2001).

[58] A. Vaziri, G. Weihs, and A. Zeilinger, Experimental two-photon, three-dimensional entanglement for quantum communication, *Phys. Rev. Lett.* **89**, 240401 (2002).

[59] A. Vaziri, J. W. Pan, T. Jennewein, G. Weihs, and A. Zeilinger, Concentration of higher dimensional entanglement: qutrits of photon orbital angular momentum, *Phys. Rev. Lett.* **91**, 227902 (2003).

[60] P. Walther, K. J. Resch, T. Rudolph, E. Schenck, H. Weinfurter, V. Vedral, M. Aspelmeyer, and A. Zeilinger, Experimental one-way quantum computing, *Nature* **434**, 169 (2005).

[61] G. Vallone, G. Donati, R. Ceccarelli, and P. Mataloni, Six-qubit two-photon hyperentangled cluster states: Characterization and application to quantum computation, *Phys. Rev. A* **81**, 052301 (2010).

[62] A. C. Dada J. Leach, G. S. Buller, M. J. Padgett, E. Andersson, Experimental high-dimensional two-photon entanglement and violations of generalized Bell inequalities, *Nat. Phys.* **7**, 677 (2011).

[63] J. Romero, D. Giovannini, S. Franke-Arnold, S. M. Barnett, and M. J. Padgett, Increasing the dimension in high-dimensional two-photon orbital angular momentum entanglement, *Phys. Rev. A* **86**, 012334 (2012).

[64] M. Krenn, M. Huber, R. Fickler, R. Lapkiewicz, S. Ramelow, and A. Zeilinger, Generation and confirmation of a (100×100)-dimensional entangled quantum system, *Proc. Natl. Acad. Sci.* **111**, 6243 (2014).

[65] A. E. Willner, H. Huang, Y. Yan, Y. Ren, N. Ahmed, G. Xie, C. Bao, L. Li, Y. Cao, Z. Zhao, J. Wang, M. P. J. Lavery, M. Tur, S. Ramachandran, A. F. Molisch, N. Ashrafi, and S. Ashrafi, Optical communications using orbital angular momentum beams, *Adv. Opt. Photon.* **7**, 66 (2015).

[66] M. Malik, M. Erhard, M. Huber, M. Krenn, R. Fickler, and A. Zeilinger, Multi-photon entanglement in high dimensions, *Nat. Photon.* **10**, 248 (2016).

[67] M. Krenn, J. Handsteiner, M. Fink, R. Fickler, R. Ursin, M. Malik, and A. Zeilinger, Twisted light transmission over 143 km, *Proc. Natl. Acad. Sci.* **113**, 13648 (2016).

[68] A. Babazadeh, M Erhard, F. Wang, M. Malik, R. Nouroozi, M. Krenn, and A. Zeilinger, High-dimensional single-photon quantum gates: concepts and experiments, *Phys. Rev. Lett.* **119**, 180510 (2017).

[69] F. Wang, M. Erhard, A. Babazadeh, M. Malik, M. Krenn, and A. Zeilinger, Generation of the complete four-dimensional Bell basis, *Optica* **4**, 1462 (2017).

[70] M. Krenn, A. Hochrainer, M. Lahiri, and A. Zeilinger, Entanglement by path identity, *Phys. Rev. Lett.* **118**, 080401 (2017).

[71] M. Erhard, R. Fickler, M. Krenn, and A. Zeilinger, Twisted photons: new quantum perspectives in high dimensions, *Light Sci. Appl.* **7**, 17146 (2018).

[72] D. Cozzolino, D. Bacco, B. Da Lio, K. Ingerslev, Y. Ding, K. Dalgaard, P. Kristensen, M. Galili, K. Rottwitt, S. Ramachandran, and L. K. Oxenløwe, Orbital angular momentum states

enabling fiber-based high-dimensional quantum communication, *Phys. Rev. Appl.* **11**, 064058 (2019).
[73] F. Brandt, M. Hiekkamäki, F. Bouchard, M. Huber, and R. Fickler, High-dimensional quantum gates using full-field spatial modes of photons, *Optica* **7**, 98 (2020).
[74] J. Kysela, M. Erhard, A. Hochrainer, M. Krenn, and A. Zeilinger, Path identity as a source of high-dimensional entanglement, *Proc. Natl. Acad. Sci.* **117**, 26118 (2020).
[75] Z. Y. Ou, and Y. J. Lu, Cavity enhanced spontaneous parametric down-conversion for the prolongation of correlation time between conjugate photons, *Phys. Rev. Lett.* **83**, 2556 (1999).
[76] Y. J. Lu, R. L. Campbell, and Z. Y. Ou, Mode-locked two-photon states, *Phys. Rev. Lett.* **91**, 163602 (2003).
[77] C. Reimer, L. Caspani, M. Clerici, M. Ferrera, M. Kues, M. Peccianti, A. Pasquazi, L. Razzari, B. E. Little, S. T. Chu, D. J. Moss, and R. Morandotti, Integrated frequency comb source of heralded single photons, *Opt. Express* **22**, 6535 (2014).
[78] C. Reimer, M. Kues, L. Caspani, B. Wetzel, P. Roztocki, M. Clerici, Y. Jestin, M. Ferrera, M. Peccianti, A. Pasquazi, B. E. Little, S. T. Chu, D. J. Moss and R. Morandotti, Cross-polarized photon-pair generation and bi-chromatically pumped optical parametric oscillation on a chip, *Nat. Commun.* **6**, 8236 (2015).
[79] Z. Xie, T. Zhong, S. Shrestha, X. Xu, J. Liang, Y. X. Gong, J.C. Bienfang, A. Restelli, J. H. Shapiro, F. N. C. Wong, and C. W. Wong, Harnessing high-dimensional hyperentanglement through a biphoton frequency comb, *Nat. Photon.* **9**, 536 (2015).
[80] C. Reimer, M. Kues, P. Roztocki, B. Wetzel, F. Grazioso, B. E. Little, S. T. Chu, T. Johnston, Y. Bromberg, L. Caspani, D. J. Moss, and R. Morandotti, Generation of multiphoton entangled quantum states by means of integrated frequency combs, *Science* **351**, 1176 (2016).
[81] J. A. Jaramillo-Villegas, P. Imany, O. D. Odele, D. E. Leaird, Z.-Y. Ou, M. Qi, and A. M. Weiner, Persistent energy-time entanglement covering multiple resonances of an on-chip biphoton frequency comb, *Optica* **4**, 655 (2017).
[82] M. Kues, C. Reimer, P. Roztocki, L. R. Cortés, S. Sciara, B. Wetzel, Y. Zhang, A. Cino, S. T. Chu, B. E. Little, D. J. Moss, L. Caspani, J. Azaña, and R. Morandotti, On-chip generation of high-dimensional entangled quantum states and their coherent control, *Nature* **546**, 622 (2017).
[83] H. Mahmudlu, R. Johanning, A. Rees, A. K. Kashi, J. P. Epping, R. Haldar, K.-J. Boller and M. Kues, Fully on-chip photonic turnkey quantum source for entangled qubit/qudit state generation, *Nat. Photon.* **17**, 518 (2023).
[84] P. Imany, J. A. Jaramillo-Villegas, O. D. Odele, K. Han, D. E. Leaird, J. M. Lukens, P. Lougovski, M. Qi, and A. M. Weiner, 50-GHz-spaced comb of high-dimensional frequency-bin entangled photons from an on-chip silicon nitride microresonator. *Opt. Express* **26**, 1825 (2018).
[85] H.-H. Lu, J. M. Lukens, N. A. Peters, B. P. Williams, A. M. Weiner, and P. Lougovski, Quantum interference and correlation control of frequency-bin qubits, *Optica* **5**, 1455 (2018).
[86] C. Reimer, S. Sciara, P. Roztocki, M. Islam, L. R. Cortés, Y. Zhang, B. Fisher, S. Loranger, R. Kashyap, A. Cino, S. T. Chu, B. E. Little, D. J. Moss, L. Caspani, W. J. Munro, J. Azaña, M. Kues, and R. Morandotti, High-dimensional one-way quantum processing implemented on $d$-level cluster states, *Nat. Phys.* **15**, 148 (2019).
[87] H.-H. Lu, J. M. Lukens, B. P. Williams, P. Imany, N. A. Peters, A. M. Weiner, and P. Lougovski, A controlled-NOT gate for frequency-bin qubits, *npj Quan. Inf.* **5**, 24 (2019).
[88] P. Imany, J. A. Jaramillo-Villegas, M. S. Alshaykh, J. M. Lukens, O. D. Odele, A. J. Moore, D. E. Leaird, M. Qi and A. M. Weiner, High-dimensional optical quantum logic in large operational spaces, *npj Quan. Inf.* **5**, 59 (2019).

[89] N. B. Lingaraju, H.-H. Lu, S. Seshadri, P. Imany, D. E. Leaird, J. M. Lukens, and A. M. Weiner, Quantum frequency combs and Hong-Ou-Mandel interferometry: the role of spectral phase coherence, *Opt. Express* **27**, 38683 (2019).

[90] R. Ikuta, R. Tani, M. Ishizaki, S. Miki, M. Yabuno, H. Terai, N. Imoto, and T. Yamamoto, Frequency-multiplexed photon pairs over 1000 modes from a quadratic nonlinear optical waveguide resonator with a singly resonant configuration, *Phys. Rev. Lett.* **123**, 193603 (2019).

[91] Y. Zhang, M. Kues, P. Roztocki, C. Reimer, B. Fischer, B. MacLellan, A. Bisianov, U. Peschel, B. E. Little, S. T. Chu, D. J. Moss, and R. Morandotti, Induced Photon Correlations Through the Overlap of Two Four-Wave Mixing Processes in Integrated Cavities, *Laser Photon. Rev.* **14**, 2000128 (2020).

[92] H.-H. Lu, K.V. Myilswamy, R. S. Bennink, S. Seshadri, M. S. Alshaykh, J. Liu, T. J. Kippenberg, D. E. Leaird, A. M. Weiner, and J. M. Lukens, Bayesian tomography of high-dimensional on-chip biphoton frequency combs with randomized measurements, *Nat. Commun*. **13**, 4338 (2022).

[93] Y. Hu, C. Reimer, A. Shams-Ansari, M. Zhang, and M. Loncar, Realization of high-dimensional frequency crystals in electro-optic microcombs, *Optica* **7**, 1189 (2020).

[94] P. Imany, N. B. Lingaraju, M. S. Alshaykh, D. E. Leaird, and A. M. Weiner, Probing quantum walks through coherent control of high-dimensionally entangled photons, *Sci. Adv.* **6**, eaba8066 (2020).

[95] C. Joshi, A. Farsi, A. Dutt, B. Y. Kim, X. Ji, Y. Zhao, A. M. Bishop, M. Lipson, and A. L. Gaeta, Frequency-Domain Quantum Interference with Correlated Photons from an Integrated Microresonator, *Phys. Rev. Lett.* **124**, 143601 (2020).

[96] N. Fabre, G. Maltese, F. Appas, S. Felicetti, A. Ketterer, A. Keller, T. Coudreau, F. Baboux, M. I. Amanti, S. Ducci, and P. Milman, Generation of a time-frequency grid state with integrated biphoton frequency combs, *Phys. Rev.* A **102**, 012607 (2020).

[97] G. Maltese, M. I. Amanti, F. Appas, G. Sinnl, A. Lemaitre, P. Milman, F. Baboux, and S. Ducci, Generation and symmetry control of quantum frequency combs, *npj Quan. Inf.* **6**, 13 (2020).

[98] N. B. Lingaraju, H.-H. Lu, S. Seshadri, D. E. Leaird, A.M. Weiner, and J. M. Lukens, Adaptive bandwidth management for entanglement distribution in quantum networks, *Optica* **8**, 329 (2021).

[99] K.-C. Chang, X. Cheng, M. C. Sarihan, A. V. Kumar, Y. S. Lee, T. Zhong, Y.-X. Gong, Z. Xie, J. H. Shapiro, F. N. C. Wong, and C. W. Wong, 648 Hilbert Space Dimensionality in a biphoton frequency comb: entanglement formation and Schmidt mode decomposition, *npj Quan. Inf.* **7**, 48 (2021).

[100] T. Yamazaki, R. Ikuta, T. Kobayashi, S. Miki, F. China, H. Terai, N. Imoto, and T. Yamamoto, Massive-mode polarization entangled biphoton frequency comb, *Sci. Rep.* **12**, 8964 (2022).

[101] L. Lu, L. Xia, Z. Chen, L. Chen, T. Yu, T. Tao, W. Ma, Y. Pan, X. Cai, Y. Lu, and S. Zhu, Three-dimensional entanglement on a silicon chip, *npj Quan. Inf.* **6**, 30 (2020).

[102] S. Slussarenko, and G. J. Pryde, Photonic quantum information processing: A concise review, *Appl. Phys. Rev.* **6**, 041303 (2019).

[103] G. B. Xavier, and G. Lima, Quantum information processing with space-division multiplexing optical fibres, *Commun. Phys.* **3**, 9 (2020).

[104] J. Wang, F. Sciarrino, A. Laing, and M. G. Thompson, Integrated photonic quantum technologies, *Nat. Photon.* **14**, 273 (2020).

[105] M. Erhard, M. Krenn, and A. Zeilinger, Advances in high-dimensional quantum entanglement, *Nat. Rev. Phys.* **2**, 365 (2020).
[106] M. Kues, C. Reimer, J. M. Lukens, W. J. Munro, A. M. Weiner, D. J. Moss, and R. Morandotti, Quantum optical microcombs, *Nat. Photon.* **13**, 170 (2019).
[107] A. Weiner, *Ultrafast optics* (Vol. 72), (John Wiley & Sons 2009).
[108] J. H. Shapiro, Coincidence dips and revivals from a Type-II optical parametric amplifier, Technical Digest of Topical Conference on Nonlinear Optics, paper FC7-1, Maui, HI, (2002).
[109] M. Rambach, W. S. Lau, S. Laibacher, V. Tamma, A. G. White, and T. J. Weinhold, Hectometer revivals of quantum interference, *Phys. Rev. Lett.* **121**, 093603 (2018).
[110] M. Scholz, L. Koch, R. Ullmann, and O. Benson, Single-mode operation of a high-brightness narrow-band single-photon source, *Appl. Phys. Lett.* **94**, 201105 (2009).
[111] K. H. Luo, V. Ansari, M. Massaro, M. Santandrea, C. Eigner, R. Ricken, H. Herrmann, and C. Silberhorn, Counter-propagating photon pair generation in a nonlinear waveguide, *Opt. Express* **28**, 3215 (2020).
[112] A. Pe'Er, B. Dayan, A. A. Friesem, and Y. Silberberg, Temporal shaping of entangled photons, *Phys. Rev. Lett.* **94**, 073601 (2005).
[113] O. Kuzucu, F. N. C. Wong, S. Kurimura, and S. Tovstonog, Joint temporal density measurements for two-photon state characterization, *Phys. Rev. Lett.* **101**, 153602 (2008).
[114] J.-P. W. MacLean, J. M. Donohue, and K. J. Resch, Direct characterization of ultrafast energy-time entangled photon pairs, *Phys. Rev. Lett.* **120**, 053601 (2018).
[115] S. Mittal, V. V. Orre, A. Restelli, R. Salem, E. A. Goldschmidt, and M. Hafezi, Temporal and spectral manipulations of correlated photons using a time lens, *Phys. Rev. A* **96**, 043807 (2017).
[116] F. A. Beduini, J. A. Zielińska, V. G. Lucivero, Y. A. de Icaza Astiz, and M. W. Mitchell, Interferometric measurement of the biphoton wave function, *Phys. Rev. Lett.* **113**, 183602 (2014).
[117] A. O. Davis, V. Thiel, M. Karpiński, and B. J. Smith, Measuring the single-photon temporal-spectral wave function, *Phys. Rev. Lett.* **121**, 083602 (2018).
[118] A. O. Davis, V. Thiel, and B. J. Smith, Measuring the quantum state of a photon pair entangled in frequency and time, *Optica* **7**, 1317 (2020).
[119] Y. C. Liu, D. J. Guo, R. Yang, C. W. Sun, J. C. Duan, Y. X. Gong, Z. Xie, and S. N. Zhu, Narrowband photonic quantum entanglement with counterpropagating domain engineering, *Photon. Res.* **9**, 1998 (2021).
[120] P. Chen, C. Shu, X. Guo, M. M. T. Loy, and S. Du, Measuring the biphoton temporal wave function with polarization-dependent and time-resolved two-photon interference, *Phys. Rev. Lett.* **114**, 010401 (2015).
[121] Z. Zhang, J. Mower, D. Englund, F. N. C. Wong, and J. H. Shapiro, Unconditional security of time-energy entanglement quantum key distribution using dual-basis interferometry, *Phys. Rev. Lett.* **112**, 120506 (2014).
[122] C. Chen, J. H. Shapiro, F. N. C. Wong, Experimental demonstration of conjugate-Franson interferometry, *Phys. Rev. Lett.* **127**, 093603 (2021).
[123] V. Averchenko, D. Sych, C. Marquardt, and G. Leuchs, Efficient generation of temporally shaped photons using non-local spectral filtering, *Phys. Rev. A* **101**, 013808 (2020).
[124] I. Gianani, M. Sbroscia, and M. Barbieri, Measuring the time-frequency properties of photon pairs: A short review, *AVS Quan. Sci.* **2**, 011701 (2020).
[125] L. Olislager, J. Cussey, A. T. Nguyen, P. Emplit, S. Massar, J. M. Merolla, and K. P. Huy, Frequency-bin entangled photons, *Phys. Rev. A* **82**, 013804 (2010).

[126] P. J. Mosley, J. S. Lundeen, B. J. Smith, P. Wasylczyk, A. B. U'Ren, C. Silberhorn, and I. A. Walmsley, Heralded generation of ultrafast single photons in pure quantum states, *Phys. Rev. Lett.* **100**, 133601 (2008).

[127] T. Gerrits, M. J. Stevens, B. Baek, B. Calkins, A. Lita, S. Glancy, E. Knill, S. W. Nam, R. P. Mirin, R. H. Hadfield, and R. S. Bennink, Generation of degenerate, factorizable, pulsed squeezed light at telecom wavelengths, *Opt. Express* **19**, 24434 (2011).

[128] B. A. Bell, G. T. Garces, and I. A. Walmsley, Diagnosing phase correlations in the joint spectrum of parametric downconversion using multi-photon emission, *Opt. Express* **28**, 34246 (2020).

[129] M. Avenhaus, A. Eckstein, P. J. Mosley, and C. Silberhorn, Fiber-assisted single-photon spectrograph, *Opt. Lett.* **34**, 2873 (2009).

[130] G. Harder, V. Ansari, B. Brecht, T. Dirmeier, C. Marquardt, and C. Silberhorn, An optimized photon pair source for quantum circuits, *Opt. Express* **21**, 13975 (2013).

[131] T. Gerrits, F. Marsili, V. B. Verma, L. K. Shalm, M. Shaw, R. P. Mirin, and S. W. Nam, Spectral correlation measurements at the Hong-Ou-Mandel interference dip, *Phys. Rev. A* **91**, 013830 (2015).

[132] M. M. Weston, H. M. Chrzanowski, S. Wollmann, A. Boston, J. Ho, L. K. Shalm, V. B. Verma, M. S. Allman, S. W. Nam, R. B. Patel, and S. Slussarenko, Efficient and pure femtosecond-pulse-length source of polarization-entangled photons, *Opt. Express* **24**, 10869 (2016).

[133] C. Chen, C. Bo, M. Y. Niu, F. Xu, Z. Zhang, J. H. Shapiro, and F. N. C. Wong, Efficient generation and characterization of spectrally factorable biphotons, *Opt. Express* **25**, 7300 (2017).

[134] V. V. Orre, E. A. Goldschmidt, A. Deshpande, A. V. Gorshkov, V. Tamma, M. Hafezi, and S. Mittal, Interference of temporally distinguishable photons using frequency-resolved detection, *Phys. Rev. Lett.* **123**, 123603 (2019).

[135] C. Chen, J. E. Heyes, K. H. Hong, M. Y. Niu, A. E. Lita, T. Gerrits, S. W. Nam, J. H. Shapiro, and F. N. C. Wong, Indistinguishable single-mode photons from spectrally engineered biphotons, *Opt. Express* **27**, 11626 (2019).

[136] F. Graffitti, P. Barrow, A. Pickston, A. M. Brańczyk, and A. Fedrizzi, Direct generation of tailored pulse-mode entanglement, *Phys. Rev. Lett.* **124**, 053603 (2020).

[137] A. Pickston, F. Graffitti, P. Barrow, C. L. Morrison, J. Ho, A. M. Brańczyk, and A. Fedrizzi, Optimised domain-engineered crystals for pure telecom photon sources, *Opt. Express* **29**, 6991 (2021).

[138] J. M. Donohue, M. Mastrovich, and K. J. Resch, Spectrally engineering photonic entanglement with a time lens, *Phys. Rev. Lett.* **117**, 243602 (2016).

[139] N. Tischler, A. Buese, L. G. Helt, M. L. Juan, N. Piro, J. Ghosh, M. J. Steel, and G. Molina-Terriza, Measurement and shaping of biphoton spectral wave functions, *Phys. Rev. Lett.* **115**, 193602 (2015).

[140] I. Jizan, B. Bell, L. G. Helt, A. C. Bedoya, C. Xiong, and B. J. Eggleton, Phase-sensitive tomography of the joint spectral amplitude of photon pair sources, *Opt. Lett.* **41**, 4803 (2016).

[141] J.-P. W. MacLean, S. Schwarz, and K. J. Resch, Reconstructing ultrafast energy-time-entangled two-photon pulses, *Phys. Rev. A* **100**, 033834 (2019).

[142] G. Triginer, M. D. Vidrighin, N. Quesada, A. Eckstein, M. Moore, W. S. Kolthammer, J. E. Sipe, and I. A. Walmsley, Understanding High-Gain Twin-Beam Sources Using Cascaded Stimulated Emission, *Phys. Rev. X* **10**, 031063 (2020).

[143] I. Gianani, Robust spectral phase reconstruction of time-frequency entangled bi-photon states, *Phys. Rev. Res.* **1**, 033165 (2019).

[144] U. A. Javid, J. Ling, J. Staffa, M. Li, Y. He, and Q. Lin, Ultra-broadband entangled photons on a nanophotonic chip, *Phys. Rev. Lett.* **127**, 183601 (2021).
[145] C. Okoth, A. Cavanna, T. Santiago-Cruz, and M. V. Chekhova, Microscale generation of entangled photons without momentum conservation, *Phys. Rev. Lett.* **123**, 263602 (2019).
[146] W. Tittel, J. Brendel, H. Zbinden, and N. Gisin, Violation of Bell inequalities by photons more than 10 km apart, *Phys. Rev. Lett.* **81**, 3563 (1998).
[147] I. Marcikic, H. De Riedmatten, W. Tittel, H. Zbinden, M. Legré, and N. Gisin, Distribution of time-bin entangled qubits over 50 km of optical fiber, *Phys. Rev. Lett.* **93**, 180502 (2004).
[148] J. D. Franson, Bell inequality for position and time, *Phys. Rev. Lett.* **62**, 2205 (1989).
[149] X. Cheng, K.-C. Chang, M. C. Sarihan, A. Mueller, M. Spiropulu, M. D. Shaw, and B. A. Korzh, A. Faraon, F. N. C. Wong, J. H. Shapiro, and C. W. Wong, High-dimensional time-frequency entanglement in a singly-filtered biphoton frequency comb, *Commun. Phys.* **6**, 278 (2023).
[150] A. Martin, T. Guerreiro, A. Tiranov, S. Designolle, F. Fröwis, N. Brunner, M. Huber, and N. Gisin, Quantifying photonic high-dimensional entanglement, *Phys. Rev. Lett.* **118**, 110501 (2017).
[151] K.-C. Chang, X. Cheng, M. C. Sarihan, and C. W. Wong, Towards optimum Franson interference recurrence in mode-locked singly-filtered biphoton frequency combs, *Photon. Res.* **11**, 1175 (2023).
[152] K.-C. Chang, X. Cheng, M. C. Sarihan, and C. W. Wong, Time-reversible and fully time-resolved ultranarrow-band biphoton frequency combs, *APL Quan.* **1**, 016106 (2024).
[153] K.-C. Chang, X. Cheng, M. C. Sarihan, F. N. C. Wong, J. H. Shapiro, and C. W. Wong, High-dimensional energy-time entanglement distribution via a biphoton frequency comb, Conference on Lasers and Electro-Optics, OSA Technical Digest (Optical Society of America, 2021), paper FF1A.7.
[154] J. Brendel, N. Gisin, W. Tittel, and H. Zbinden, Pulsed energy-time entangled twin-photon source for quantum communication, *Phys. Rev. Lett.* **82**, 2594 (1999).
[155] R. T. Thew, S. Tanzilli, W. Tittel, H. Zbinden, and N. Gisin, Experimental investigation of the robustness of partially entangled qubits over 11 km, *Phys. Rev. A* **66**, 062304 (2002).
[156] H. De Riedmatten, I. Marcikic, V. Scarani, W. Tittel, H. Zbinden, and N. Gisin, Tailoring photonic entanglement in high-dimensional Hilbert spaces, *Phys. Rev. A* **69**, 050304 (2004).
[157] T. Ikuta, and H. Takesue, Implementation of quantum state tomography for time-bin qudits, *New J. Phys.* **19**, 013039 (2017).
[158] T. Brougham, S. M. Barnett, K. T. McCusker, P. G. Kwiat, and D. J. Gauthier, Security of high-dimensional quantum key distribution protocols using Franson interferometers, *J. Phys. B: At., Mol. Opt. Phys.* **46**, 104010 (2013).
[159] N. Friis, G. Vitagliano, M. Malik, and M. Huber, Entanglement certification from theory to experiment, *Nat. Rev. Phys.* **1**, 72 (2019).
[160] D. Giovannini, J. Romero, J. Leach, A. Dudley, A. Forbes, and M. J. Padgett, Characterization of high-dimensional entangled systems via mutually unbiased measurements, *Phys. Rev. Lett.* **110**, 143601 (2013).
[161] P. J. Coles, M. Berta, M. Tomamichel, and S. Wehner, Entropic uncertainty relations and their applications, *Rev. Mod. Phys.* **89**, 015002 (2017).
[162] Y. Ding, D. Bacco, K. Dalgaard, X. Cai, X. Zhou, K. Rottwitt, and L. K. Oxenløwe, High-dimensional quantum key distribution based on multicore fiber using silicon photonic integrated circuits, *npj Quan. Inf.* **3**, 25 (2017).

[163] J. Bavaresco, N. H. Valencia, C. Klöckl, M. Pivoluska, P. Erker, N. Friis, M. Malik, and M. Huber, Measurements in two bases are sufficient for certifying high-dimensional entanglement, *Nat. Phys.* **14**, 1032 (2018).

[164] M. Lucamarini, K. A. Patel, J. F. Dynes, B. Fröhlich, A. W. Sharpe, A. R. Dixon, Z. L. Yuan, R. V. Penty, and A. J. Shields, Efficient decoy-state quantum key distribution with quantified security, *Opt. Express* **21** 24550 (2013).

[165] J. Mower, Z. Zhang, P. Desjardins, C. Lee, J. H. Shapiro, and D. Englund, High-dimensional quantum key distribution using dispersive optics, *Phys. Rev. A* **87**, 062322 (2013).

[166] C. Lee, Z. Zhang, G. R. Steinbrecher, H. Zhou, J. Mower, T. Zhong, L. Wang, X. Hu, R. D. Horansky, V. B. Verma, A. E. Lita, R. P. Mirin, F. Marsili, M. D. Shaw, S. W. Nam, G. W. Wornell, F. N. C. Wong, J. H. Shapiro, and D. Englund, Entanglement-based quantum communication secured by non-local dispersion cancellation, *Phys. Rev. A* **90**, 062331 (2014).

[167] T. Ikuta, S. Akibue, Y. Yonezu, T. Honjo, H. Takesue, and K. Inoue, Scalable implementation of ($d$+1) mutually unbiased bases for $d$-dimensional quantum key distribution, *Phys. Rev. Res.* **4**, L042007 (2022).

[168] J. M. Lukens, N. T. Islam, C. C. W. Lim, and D. J. Gauthier, Reconfigurable generation and measurement of mutually unbiased bases for time-bin qudits, *Appl. Phys. Lett.* **112**, 111102 (2018).

[169] A. K. Ekert, Quantum cryptography based on Bell's theorem, *Phys. Rev. Lett.* **67**, 661 (1991).

[170] F. Xu, X. Ma, Q. Zhang, H. K. Lo, and J. W. Pan, Secure quantum key distribution with realistic devices, *Rev. Mod. Phys.* **92**, 025002 (2020).

[171] J. M. Lukens, and P. Lougovski, Frequency-encoded photonic qubits for scalable quantum information processing, *Optica* **4**, 8 (2017).

[172] A. M. Weiner, Femtosecond pulse shaping using spatial light modulators, *Rev. Sci. Inst.* **71**, 1929 (2000).

[173] M. Scholz, L. Koch, and O. Benson, Statistics of narrow-band single photons for quantum memories generated by ultrabright cavity-enhanced parametric down-conversion, *Phys. Rev. Lett.* **102**, 063603 (2009).

[174] C. K. Law, I. A. Walmsley, and J. H. Eberly, Continuous frequency entanglement: effective finite Hilbert space and entropy control, *Phys. Rev. Lett.* **84**, 5304 (2000).

[175] A. Christ, K. Laiho, A. Eckstein, K. N. Cassemiro, and C. Silberhorn, Probing multimode squeezing with correlation functions, *New J. Phys.* **13**, 033027 (2011).

[176] M. Liscidini, and J. E. Sipe, Scalable and efficient source of entangled frequency bins, *Opt. Lett.* **44**, 2625 (2019).

[177] C. Wang, M. Zhang, X. Chen, M. Bertrand, A. Shams-Ansari, S. Chandrasekhar, P. Winzer, and M. Lončar, Integrated lithium niobate electro-optic modulators operating at CMOS-compatible voltages, *Nature* **562**, 101 (2018).

[178] P. Kharel, C. Reimer, K. Luke, L. He, and M. Zhang, Breaking voltage-bandwidth limits in integrated lithium niobate modulators using micro-structured electrodes, *Optica* **8**, 357 (2021).

[179] H.-H. Lu, N. A. Peters, A. M. Weiner and J. M. Lukens, Characterization of Quantum Frequency Processors, *IEEE J. Sel. Top. Quan. Electron.* **29**, 1 (2023).

[180] H.-H. Lu, M. Liscidini, A. L. Gaeta, A. M. Weiner, and J. M. Lukens, Frequency-bin photonic quantum information, *Optica* **10**, 1655 (2023).

[181] J. T. Barreiro, N. K. Langford, N. A. Peters, and P. G. Kwiat, Generation of hyperentangled photon pairs, *Phys. Rev. Lett.* **95**, 260501 (2005).

[182] M. Erhard, M. Malik, M. Krenn, and A. Zeilinger, Experimental Greenberger-horne-zeilinger entanglement beyond qubits, *Nat. Photon.* **12**, 759 (2018).

[183] A. Einstein, B. Podolsky, and N. Rosen, Can quantum-mechanical description of physical reality be considered complete? *Phys. Rev.* **47**, 777 (1935).

[184] N. D. Mermin, Extreme quantum entanglement in a superposition of macroscopically distinct states, *Phys. Rev. Lett.* **65**, 1838 (1990).

[185] J. L. O'Brien, G. J. Pryde, A. G. White, T. C. Ralph, and D. Branning, Demonstration of an all-optical quantum controlled-NOT gate, *Nature* **426**, 264 (2003).

[186] M. Fiorentino and F. N. C. Wong, Deterministic controlled-NOT gate for single-photon two-qubit quantum logic, *Phys. Rev. Lett.* **93**, 070502 (2004).

[187] M. Fiorentino, T. Kim, and F. N. C. Wong, Single-photon two-qubit SWAP gate for entanglement manipulation, *Phys. Rev. A* **72**, 012318 (2005).

[188] A. Politi, M. J. Cryan, J. G. Rarity, S. Yu, and J. L. O'brien, Silica-on-silicon waveguide quantum circuits, *Science* **320**, 646 (2008).

[189] J. Carolan, C. Harrold, C. Sparrow, E. Martín-López, N. J. Russell, J. W. Silverstone, P. J. Shadbolt, N. Matsuda, M. Oguma, M. Itoh, G. D. Marshall, M. G. Thompson, J. C. F. Matthews, T. Hashimoto, J. L. O'Brien, and A. Laing, Universal linear optics, *Science* **349**, 711 (2015).

[190] X. Cheng, K.-C. Chang, Z. Xie, M. C. Sarihan, Y. S. Lee, Y. Li, X. Xu, A. V. Kumar, S. Kocaman, M. Yu, P. G.-Q. Lo, D.-L. Kwong, J. H. Shapiro, F. N. C. Wong, and C. W. Wong, A chip-scale polarization-spatial-momentum quantum SWAP gate in silicon nanophotonics, *Nat. Photon.* **17**, 656 (2023).

[191] B. P. Williams, R. J. Sadlier, and T. S. Humble, Superdense coding over optical fiber links with complete Bell-state measurements, *Phys. Rev. Lett.* **118**, 050501 (2017).

[192] F. Steinlechner, S. Ecker, M. Fink, B. Liu, J. Bavaresco, M. Huber, T. Scheidl, and R. Ursin, Distribution of high-dimensional entanglement via an intra-city free-space link, *Nat. Commun.* **8**, 15971 (2017).

[193] P. Vergyris, F. Mazeas, E. Gouzien, L. Labonté, O. Alibart, S. Tanzilli, and F. Kaiser, Fibre based hyperentanglement generation for dense wavelength division multiplexing, *Quan. Sci. Technol.* **4**, 045007 (2019).

[194] A. Tiranov, J. Lavoie, A. Ferrier, P. Goldner, V. B. Verma, S. W. Nam, R. P. Mirin, nA. E. Lita, F. Marsili, H. Herrmann, and C. Silberhorn, Storage of hyperentanglement in a solid-state quantum memory, *Optica* **2**, 279 (2015).

[195] J. C. Chapman, T. M. Graham, C. K. Zeitler, H. J. Bernstein, and P. G. Kwiat, Time-bin and polarization superdense teleportation for space applications, *Phys. Rev. Appl.* **14**, 014044 (2020).

[196] S. Lloyd, J. H. Shapiro, and F. N. C. Wong, Quantum magic bullets by means of entanglement, *J. Opt. Soc. Am. B* **19**, 312 (2002).

[197] H. Wang, T. Horikiri, and T. Kobayashi, Polarization-entangled mode-locked photons from cavity-enhanced spontaneous parametric down-conversion, *Phys. Rev. A* **70**, 043804 (2004).

[198] C. E. Kuklewicz, F. N. C. Wong, and J. H. Shapiro, Time-bin-modulated biphotons from cavity-enhanced down-conversion, *Phys. Rev. Lett.* **97**, 223601 (2006).

[199] F. Wolfgramm, Y. A. de Icaza Astiz, F. A. Beduini, A. Cere, and M. W. Mitchell, Atom-resonant heralded single photons by interaction-free measurement, *Phys. Rev. Lett.* **106**, 053602 (2011).

[200] J. Fekete, D. Rieländer, M. Cristiani, and H. de Riedmatten, Ultranarrow-band photon-pair source compatible with solid state quantum memories and telecommunication networks, *Phys. Rev. Lett.* **110**, 220502 (2013).

[201] D. Rieländer, K. Kutluer, P. M. Ledingham, M. Gündoğan, J. Fekete, M. Mazzera, and H. De Riedmatten, Quantum storage of heralded single photons in a praseodymium-doped crystal, *Phys. Rev. Lett.* **112**, 040504 (2014).

[202] A. Seri, A. Lenhard, D. Rieländer, M. Gündoğan, P. M. Ledingham, M. Mazzera, and H. De Riedmatten, Quantum correlations between single telecom photons and a multimode on-demand solid-state quantum memory, *Phys. Rev. X* **7**, 021028 (2017).

[203] A. Seri, G. Corrielli, D. Lago-Rivera, A. Lenhard, H. de Riedmatten, R. Osellame, and M. Mazzera, Laser-written integrated platform for quantum storage of heralded single photons, *Optica* **5**, 934 (2018).

[204] A. Seri, D. Lago-Rivera, A. Lenhard, G. Corrielli, R. Osellame, M. Mazzera, and H. de Riedmatten, Quantum storage of frequency-multiplexed heralded single photons, *Phys. Rev. Lett.* **123**, 080502 (2019).

[205] D. Lago-Rivera, S. Grandi, J. V. Rakonjac, A. Seri and H. de Riedmatten, Telecom-heralded entanglement between multimode solid-state quantum memories, *Nature* **594**, 37 (2021).

[206] J. H. Shapiro, C. E. Kuklewicz, and F. N. C. Wong, Quantum signatures from singly-resonant and doubly-resonant parametric amplifiers. In Nonlinear Optics: Materials, Fundamentals and Applications (p.TuA5), Optical Society of America (2004).

[207] O. Slattery, L. Ma, K. Zong, and X. Tang, Background and review of cavity-enhanced spontaneous parametric down-conversion, *J. Res. Natl. Inst. Stan. Technol.* **124**, 1 (2019).

[208] . N. Sangouard, C. Simon, H. De Riedmatten, and N. Gisin, Quantum repeaters based on atomic ensembles and linear optics, *Rev. Mod. Phys.* **83**, 33 (2011).

[209] M. Rambach, A. Nikolova, T. J. Weinhold, and A. G. White, Sub-megahertz linewidth single photon source, *APL Photon.* **1**, 096101 (2016).

[210] C. Müller, A. Ahlrichs, and O. Benson, General and complete description of temporal photon correlations in cavity-enhanced spontaneous parametric down-conversion, *Phys. Rev. A* **102**, 053504 (2020).

[211] M. Scholz, F. Wolfgramm, U. Herzog, and O. Benson, Narrow-band single photons from a single-resonant optical parametric oscillator far below threshold, *Appl. Phys. Lett.* **91**, 191104 (2007).

[212] A. Lenhard, M. Bock, C. Becher, S. Kucera, J. Brito, P. Eich, P. Müller, and J. Eschner, Telecom-heralded single-photon absorption by a single atom, *Phys. Rev. A* **92**, 063827 (2015).

[213] E. Pomarico, B. Sanguinetti, N. Gisin, R. Thew, H. Zbinden, G. Schreiber, A. Thomas, and W. Sohler, Waveguide-based OPO source of entangled photon pairs, *New J. Phys.* **11**, 113042 (2009).

[214] A. Ahlrichs, and O. Benson, Bright source of indistinguishable photons based on cavity-enhanced parametric down-conversion utilizing the cluster effect, *Appl. Phys. Lett.* **108**, 021111 (2016).

[215] C. S. Chuu, G. Y. Yin, and S. E. Harris, A miniature ultrabright source of temporally long, narrowband biphotons, *Appl. Phys. Lett.* **101**, 051108 (2012).

[216] D. Rieländer, A. Lenhard, M. Mazzera, and H. De Riedmatten, Cavity enhanced telecom heralded single photons for spin-wave solid state quantum memories, *New J. Phys.* **18**, 123013 (2016).

[217] P. J. Tsai, and Y. C. Chen, Ultrabright, narrow-band photon-pair source for atomic quantum memories, *Quan. Sci. Technol.* **3**, 034005 (2018).

[218] A. Moqanaki, F. Massa, and P. Walther, Novel single-mode narrow-band photon source of high brightness tuned to cesium D2 line, *APL Photon.* **4**, 090804 (2019).

[219] R. Kumar, J. R. Ong, M. Savanier, and S. Mookherjea, Controlling the spectrum of photons generated on a silicon nanophotonic chip, *Nat. Commun.* **5**, 5489 (2014).
[220] R. Kumar, M. Savanier, J. R. Ong, and S. Mookherjea, Entanglement measurement of a coupled silicon microring photon pair source, *Opt. Express* **23**, 19318 (2015).
[221] E. Saglamyurek, N. Sinclair, J. Jin, J. A. Slater, D. Oblak, F. Bussières, M. George, R. Ricken, W. Sohler, and W. Tittel, Broadband waveguide quantum memory for entangled photons, *Nature* **469**, 512 (2011).
[222] C. Clausen, I. Usmani, F. Bussieres, N. Sangouard, M. Afzelius, H. de Riedmatten, and N. Gisin, Quantum storage of photonic entanglement in a crystal, *Nature* **469**, 508 (2011).
[223] E. Saglamyurek, J. Jin, V. B. Verma, M. D. Shaw, F. Marsili, S. W. Nam, D. Oblak, and W. Tittel, Quantum storage of entangled telecom-wavelength photons in an erbium-doped optical fibre, *Nat. Photon.* **9**, 83 (2015).
[224] T. Kim, M. Fiorentino, and F. N. C. Wong, Phase-stable source of polarization-entangled photons using a polarization Sagnac interferometer, *Phys. Rev. A* **73,** 012316 (2006).
[225] O. Kuzucu, and F. N. C. Wong, Pulsed Sagnac source of narrow-band polarization-entangled photons, *Phys. Rev. A* **77**, 032314 (2008).
[226] T. Honjo, H. Takesue, H. Kamada, Y. Nishida, O. Tadanaga, M. Asobe, and K. Inoue,. Long-distance distribution of time-bin entangled photon pairs over 100 km using frequency up-conversion detectors, *Opt. Express* **15**, 13957 (2007).
[227] J. F. Dynes, H. Takesue, Z. L. Yuan, A. W. Sharpe, K. Harada, T. Honjo, H. Kamada, O. Tadanaga, Y. Nishida, M. Asobe, and A. J. Shields, Efficient entanglement distribution over 200 kilometers, *Opt. Express* **17**, 11440 (2009).
[228] T. Inagaki, N. Matsuda, O. Tadanaga, M. Asobe, and H. Takesue, Entanglement distribution over 300 km of fiber, *Opt. Express* **21**, 23241 (2013).
[229] P. Toliver, J. M. Dailey, A. Agarwal, and N. A. Peters, Continuously active interferometer stabilization and control for time-bin entanglement distribution, *Opt. Express* **23**, 4135 (2015).
[230] T. Ikuta, and H. Takesue, Four-dimensional entanglement distribution over 100 km, *Sci. Rep.* **8**, 1 (2018).
[231] H. Takesue, K. I. Harada, K. Tamaki, H. Fukuda, T. Tsuchizawa, T. Watanabe, K. Yamada, and S. I. Itabashi, Long-distance entanglement-based quantum key distribution experiment using practical detectors, *Opt. Express* **18**, 16777 (2010).
[232] M. Y. Niu, F. Xu, J. H. Shapiro, and F. Furrer, Finite-key analysis for time-energy high-dimensional quantum key distribution, *Phys. Rev. A* **94**, 052323 (2016).
[233] J. Jin, J. P. Bourgoin, R. Tannous, S. Agne, C. J. Pugh, K. B. Kuntz, B. L. Higgins, and T. Jennewein, Genuine time-bin-encoded quantum key distribution over a turbulent depolarizing free-space channel, *Opt. Express* **27**, 37214 (2019).
[234] J. D. Franson, Non-local cancellation of dispersion, *Phys. Rev. A* **45**, 3126 (1992).
[235] J. D. Franson, Nonclassical nature of dispersion cancellation and non-local interferometry, *Phys. Rev. A* **80**, 032119 (2009).
[236] K.-C. Chang, M. C. Sarihan, X. Cheng, Z. Zhang, and C. W. Wong, Large-alphabet time-bin quantum key distribution and Einstein-Podolsky-Rosen steering via dispersive optics, *Quan. Sci. Technol.* **9**, 015018 (2024).
[237] X. Liu, X. Yao, R. Xue, H. Wang, H. Li, Z. Wang, L. You, X. Feng, F. Liu, K. Cui, and Y. Huang, An entanglement-based quantum network based on symmetric dispersive optics quantum key distribution, *APL Photon.* **5**, 076104 (2020).

[238] D. Aktas, B. Fedrici, F. Kaiser, T. Lunghi, L. Labonté, and S. Tanzilli, Entanglement distribution over 150 km in wavelength division multiplexed channels for quantum cryptography, *Laser Photon. Rev.* **10**, 451 (2016).
[239] S. Wengerowsky, S. Koduru, F. Steinlechner, H. Hubel, and R. Ursin, An entanglement-based wavelength-multiplexed quantum communication network, *Nature* **564**, 225 (2018).
[240] S. K. Joshi, D. Aktas, S. Wengerowsky, M. Lončarić, S. P. Neumann, B. Liu, T. Scheidl, G. C. Lorenzo, Ž. Samec, L. Kling, A. Qiu, M. Razavi, M. Stipcevic, J. G. Rarity and R. Ursin, A trusted node-free eight-user metropolitan quantum communication network, *Sci. Adv.* **6**, eaba0959 (2020).
[241] J.-H. Kim, J.-W. Chae, Y.-C. Jeong, and Y.-H. Kim, Quantum communication with time-bin entanglement over a wavelength-multiplexed fiber network, *APL Photon.* **7**, 016106 (2022).
[242] R. Fujimoto, T. Yamazaki, T. Kobayashi, S. Miki, F. China, H. Terai, R. Ikuta, and T. Yamamoto, Entanglement distribution using a biphoton frequency comb compatible with DWDM technology, *Opt. Express* **30**, 36711 (2022).
[243] K. Niizeki, D. Yoshida, K. Ito, I. Nakamura, N. Takei, K. Okamura, M. Y. Zheng, X. P. Xie, and T. Horikiri, Two-photon comb with wavelength conversion and 20-km distribution for quantum communication, *Commun. Phys.* **3**, 138 (2020).
[244] M. Bourennane, A. Karlsson, and G. Björk, Quantum key distribution using multilevel encoding. *Phys. Rev. A* **64**, 012306 (2001).
[245] D. Cozzolino, B. Da Lio, D. Bacco, and L. K. Oxenløwe, High-Dimensional Quantum Communication: Benefits, Progress, and Future Challenges, *Adv. Quan. Technol.* **2**, 1900038 (2019).
[246] S. Pirandola, U. L. Andersen, L. Banchi, M. Berta, D. Bunandar, R. Colbeck, D. Englund, T. Gehring, C. Lupo, C. Ottaviani, J. L. Pereira, M. Razavi, J. Shamsul Shaari, M. Tomamichel, V. C. Usenko, G. Vallone, P. Villoresi, and P. Wallden, Advances in quantum cryptography, *Adv. Opt. Photon.* **12**, 1012 (2020).
[247] T. K. Paraïso, R. I. Woodward, D. G. Marangon, V. Lovic, Z. Yuan, and A. J. Shields, Advanced Laser Technology for Quantum Communications (Tutorial Review), *Adv. Quan. Technol.* **4**, 2100062 (2021).
[248] F. Bouchard, D. England, P. J. Bustard, K. L. Fenwick, E. Karimi, K. Heshami, and B. Sussman, Achieving ultimate noise tolerance in quantum communication, *Phys. Rev. Appl.* **15**, 024027 (2021).
[249] S.-K. Liao, W.-Q. Cai, W.-Y. Liu, L. Zhang, Y. Li, J.-G. Ren, J. Yin, Q. Shen, Y. Cao, Z.-P. Li, F.-Z. Li, X.-W. Chen, L.-H. Sun, J.-J. Jia, J.-C. Wu, X.-J. Jiang, J.-F. Wang, Y.-M. Huang, Q. Wang, Y.-L. Zhou, L. Deng, T. Xi, L. Ma, T. Hu, Q. Zhang, Y.-A. Chen, N.-L. Liu, X.-B. Wang, Z.-C. Zhu, C.-Y. Lu, R. Shu, C.-Z. Peng, J.-Y. Wang, J.-W. Pan, Satellite-to-ground quantum key distribution, *Nature* **549**, 43 (2017).
[250] S.-K. Liao, W.-Q. Cai, J. Handsteiner, B. Liu, J. Yin, L. Zhang, D. Rauch, M. Fink, J.-G. Ren, W.-Y. Liu, Y. Li, Q. Shen, Y. Cao, F.-Z. Li, J.-F. Wang, Y.-M. Huang, L. Deng, T. Xi, L. Ma, T. Hu, L. Li, N.-L. Liu, F. Koid, P. Wang, Y.-A. Chen, X.-B. Wang, M. Steindorfer, G. Kirchner, C.-Y. Lu, R. Shu, R. Ursin, T. Scheid, C.-Z. Peng, J.-Y. Wang, A. Zeilinger, and J.-W. Pan, Satellite-relayed intercontinental quantum network, *Phys. Rev. Lett.* **120**, 030501 (2018).
[251] J. Yin, Y.-H. Li, S.-K. Liao, M. Yang, Y. Cao, L. Zhang, J.-G. Ren, W.-Q. Cai, W.-Y. Liu, S.-L. Li, R. Shu, Y.-M. Huang, L. Deng, L. Li, Q. Zhang, N.-L. Liu, Y.-A. Chen, C.-Y. Lu, X.-B. Wang, F. Xu, J.-Y. Wang, C.-Z. Peng, A. K. Ekert and J.-W. Pan, Entanglement-based secure quantum cryptography over 1,120 kilometres, *Nature* **582**, 501 (2020).

[252] Y.-A. Chen, Q. Zhang, T.-Y. Chen, W.-Q. Cai, S.-K. Liao, J. Zhang, K. Chen, J. Yin, J.-G. Ren, Z. Chen, S.-L. Han, Q. Yu, K. Liang, F. Zhou, X. Yuan, M.-S. Zhao, T.-Y. Wang, X. Jiang, L. Zhang, W.-Y. Liu, Y. Li, Q. Shen, Y. Cao, C.-Y. Lu, R. Shu, J.-Y. Wang, L. Li, N.-L. Liu, F. Xu, X.-B. Wang, C.-Z. Peng, and J.-W. Pan, An integrated space-to-ground quantum communication network over 4,600 kilometres, *Nature* **589**, 214 (2021).
[253] A. Muthukrishnan, and C. R. Stroud Jr, Multivalued logic gates for quantum computation, *Phys. Rev. A* **62**, 052309 (2000).
[254] M. Luo, and X. Wang, Universal quantum computation with qudits, *Sci. China Phys. Mech. Astr.* **57**, 1712 (2014).
[255] U. A. Javid, R. Lopez-Rios, J. Ling, A. Graf, J. Staffa, and Q. Lin, Chip-scale simulations in a quantum-correlated synthetic space, *Nat. Photon.* **17**, 883 (2023).
[256] S. S. Bullock, D. P. O'Leary, and G. K. Brennen, Asymptotically optimal quantum circuits for *d*-level systems, *Phys. Rev. Lett.* **94**, 230502 (2005).
[257] M. Y. Niu, I. L. Chuang, and J. H. Shapiro, Qudit-basis universal quantum computation using $\chi^{(2)}$ interactions, *Phys. Rev. Lett.* **120**, 160502 (2018).
[258] R. N. Alexander, P. Wang, N. Sridhar, M. Chen, O. Pfister, and N. C. Menicucci, One-way quantum computing with arbitrarily large time-frequency continuous-variable cluster states from a single optical parametric oscillator, *Phys. Rev. A* **94**, 032327 (2016).
[259] T. Yamazaki, T. Arizono, T. Kobayashi, R. Ikuta, and T. Yamamoto, Linear optical quantum computation with frequency-comb qubits and passive devices, *Phys. Rev. Lett.* **130**, 200602 (2023).
[260] H. J. Briegel, and R. Raussendorf, Persistent entanglement in arrays of interacting particles, *Phys. Rev. Lett.* **86**, 910 (2001).
[261] R. Raussendorf, and H. J. Briegel, A one-way quantum computer, *Phys. Rev. Lett.* **86**, 5188 (2001).
[262] M. A. Nielsen, Optical quantum computation using cluster states, *Phys. Rev. Lett.* **93**, 040503 (2004).
[263] Y. Wang, Z. Hu, B. C. Sanders, and S. Kais, Qudits and high-dimensional quantum computing, *Fron. Phys.* **8**, 479 (2020).
[264] D. Awschalom, K. K. Berggren, H. Bernien, S. Bhave, L. D. Carr, P. Davids, S. E. Economou, D. Englund, A. Faraon, M. Fejer, S. Guha, M. V. Gustafsson, E. Hu, L. Jiang, J. Kim, B. Korzh, P. Kumar, P. G. Kwiat, M. Loncar, M. D. Lukin, D. A. B. Miller, C. Monroe, S. W. Nam, P. Narang, J. S. Orcutt, M. G. Raymer, A. H. Safavi-Naeini, M. Spiropulu, K. Srinivasan, S. Sun, J. Vučković, E. Waks, R. Walsworth, A. M. Weiner, and Z. Zhang, Development of quantum interconnects (quics) for next-generation information technologies, *PRX Quan.* **2**, 017002 (2021).
[265] P. Kumar, Quantum frequency conversion, *Opt. Lett.* **15**, 1476 (1990).
[266] M. Bock, P. Eich, S. Kucera, M. Kreis, A. Lenhard, C. Becher, and J. Eschner, High-fidelity entanglement between a trapped ion and a telecom photon via quantum frequency conversion, *Nat. Commun.* **9**, 1998 (2018).
[267] A. P. Higginbotham, P. S. Burns, M. D. Urmey, R. W. Peterson, N. S. Kampel, B. M. Brubaker, G. Smith, K. W. Lehnert and C. A. Regal, Harnessing electro-optic correlations in an efficient mechanical converter, *Nat. Phys.* **14**, 1038 (2018).
[268] Z. Xie, K.-H. Lou, K.-C. Chang, N. C. Panoiu, H. Herrmann, C. Silberhorn, and C. W. Wong, Efficient C-band single-photon upconversion with chip-scale Ti-indiffused pp-LiNbO3 waveguides, *Appl. Opt.* **58**, 5910 (2019).
[269] N. Lauk, N. Sinclair, S. Barzanjeh, J. P. Covey, M. Saffman, M. Spiropulu, and C. Simon, Perspectives on quantum transduction, *Quan. Sci. Technol.* **5**, 020501 (2020).

[270] J. G. Bartholomew, J. Rochman, T. Xie, J. M. Kindem, A. Ruskuc, I. Craiciu, M. Lei, and A. Faraon, On-chip coherent microwave-to-optical transduction mediated by ytterbium in YVO$_4$, *Nat. Commun.* **11**, 3266 (2020).

[271] M. Mirhosseini, A. Sipahigil, M. Kalaee, and O. Painter, Superconducting qubit to optical photon transduction, *Nature* **588**, 599 (2020).

[272] K. Sanaka, K. Kawahara, and T. Kuga, Experimental probabilistic manipulation of down-converted photon pairs using unbalanced interferometers, *Phys. Rev. A* **66**, 040301 (2002).

[273] J. M. Donohue, M. Agnew, J. Lavoie, and K. J. Resch, Coherent ultrafast measurement of time-bin encoded photons, *Phys. Rev. Lett.* **111**, 153602 (2013).

[274] R. Fickler, R. Lapkiewicz, M. Huber, M. P. J. Lavery, M. J. Padgett, and A. Zeilinger, Interface between path and orbital angular momentum entanglement for high-dimensional photonic quantum information, *Nat. Commun.* **5**, 4502 (2014).

[275] C. Kupchak, P. J. Bustard, K. Heshami, J. Erskine, M. Spanner, D. G. England, and B. J. Sussman, Time-bin-to-polarization conversion of ultrafast photonic qubits, *Phys. Rev. A* **96**, 053812 (2017).

[276] C. L. Degen, F. Reinhard, and P. Cappellaro, Quantum sensing, *Rev. Mod. Phys.* **89**, 035002 (2017).

[277] E. Altman, K. R. Brown, G. Carleo, L. D. Carr, E. Demler, C. Chin, B. DeMarco, S. E. Economou, M. A. Eriksson, K.-M. C. Fu, M. Greiner, K. R.A. Hazzard, R. G. Hulet, A. J. Kollár, B. L. Lev, M. D. Lukin, R. Ma, X. Mi, S. Misra, C. Monroe, K. Murch, Z. Nazario, K.-K. Ni, A. C. Potter, Pe. Roushan, M. Saffman, M. Schleier-Smith, I. Siddiqi, R. Simmonds, M. Singh, I. B. Spielman, K. Temme, D. S. Weiss, J. Vučković, V. Vuletic, J. Ye, and M. Zwierlein, Quantum simulators: Architectures and opportunities, *PRX Quan.* **2**, 017003 (2021).

[278] Y. Alexeev, D. Bacon, K. R. Brown, R. Calderbank, L. D. Carr, F. T. Chong, B. DeMarco, D. Englund, E. Farhi, B. Fefferman, A. V. Gorshkov, A. Houck, J. Kim, S. Kimmel, M. Lange, S. Lloyd, M. D. Lukin, D. Maslov, P. Maunz, C. Monroe, J. Preskill, M. Roetteler, M. J. Savage, and J. Thompson, Quantum computer systems for scientific discovery, *PRX Quan.* **2**, 017001 (2021).

[279] H. H. Lu, J. M. Lukens, N. A. Peters, O. D. Odele, D. E. Leaird, A. M. Weiner, and P. Lougovski, Electro-optic frequency beam splitters and tritters for high-fidelity photonic quantum information processing, *Phys. Rev. Lett.* **120**, 030502 (2018).

[280] H. H. Lu, E. M. Simmerman, P. Lougovski, A. M. Weiner, and J. M. Lukens, Fully arbitrary control of frequency-bin qubits, *Phys. Rev. Lett.* **125**, 120503 (2020).

[281] S. Seshadri, H. H. Lu, D. E. Leaird, A. M. Weiner, and J. M. Lukens, Complete frequency-bin Bell basis synthesizer, *Phys. Rev. Lett.* **129**, 230505 (2022).

[282] A. Tiranov, S. Designolle, E. Z. Cruzeiro, J. Lavoie, N. Brunner, M. Afzelius, M. Huber, and N. Gisin, Quantification of multidimensional entanglement stored in a crystal, *Phys. Rev. A* **96**, 040303 (2017).

[283] S. H. Wei, B. Jing, X. Y. Zhang, J. Y. Liao, C. Z. Yuan, B. Y. Fan, C. Lyu, D. L. Zhou, Y. Wang, G. W. Deng, H. Z. Song, D. Oblak, G. C. Guo, and Q. Zhou, Towards real-world quantum networks: a review, *Laser Photon. Rev.* **16**, 2100219 (2022).

[284] X. Zhang, B. Zhang, S. Wei, H. Li, J. Liao, C. Li, G. Deng, Y. Wang, H. Song, L. You, and B. Jing, F. Chen, G. C. Guo, and Q. Zhou, Telecom-band–integrated multimode photonic quantum memory, *Sci. Adv.* **9**, eadf4587 (2023).

[285] M. H. Jiang, W. Xue, Q. He, Y. Y. An, X. Zheng, W. J. Xu, Y. B. Xie, Y. Lu, S. Zhu, and X. S. Ma, Quantum storage of entangled photons at telecom wavelengths in a crystal, *Nat. Commun.* **14**, 6995 (2023).

[286] Y. Lei, F. K. Asadi, T. Zhong, A. Kuzmich, C. Simon, and M. Hosseini, Quantum optical memory for entanglement distribution, *Optica* **10**, 1511 (2023).

[287] S. H. Wei, B. Jing, X. Y. Zhang, J. Y. Liao, H. Li, L. X. You, Z. Wang, Y. Wang, G. W. Deng, H. Z. Song, and D. Oblak, G. C. Guo, and Q. Zhou, Quantum storage of 1650 modes of single photons at telecom wavelength, *npj Quan. Inf.* **10**, 19 (2024).

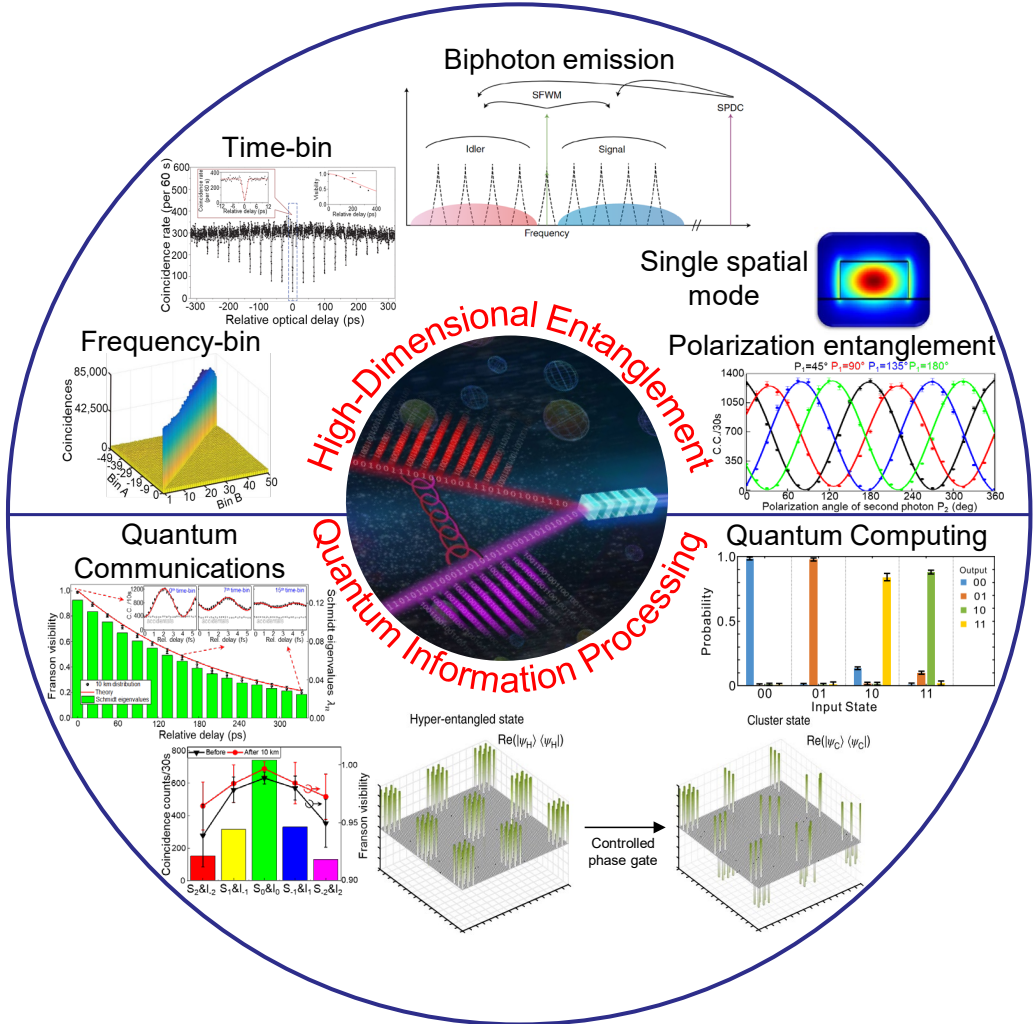

**FIG. 1 Properties and applications of mode-locked quantum frequency combs.** Mode-locked quantum frequency combs feature the biphoton emission, combined DoFs in a single spatial mode, and a variety of high-dimensional entangled states can be generated to pioneer emerging research fields.

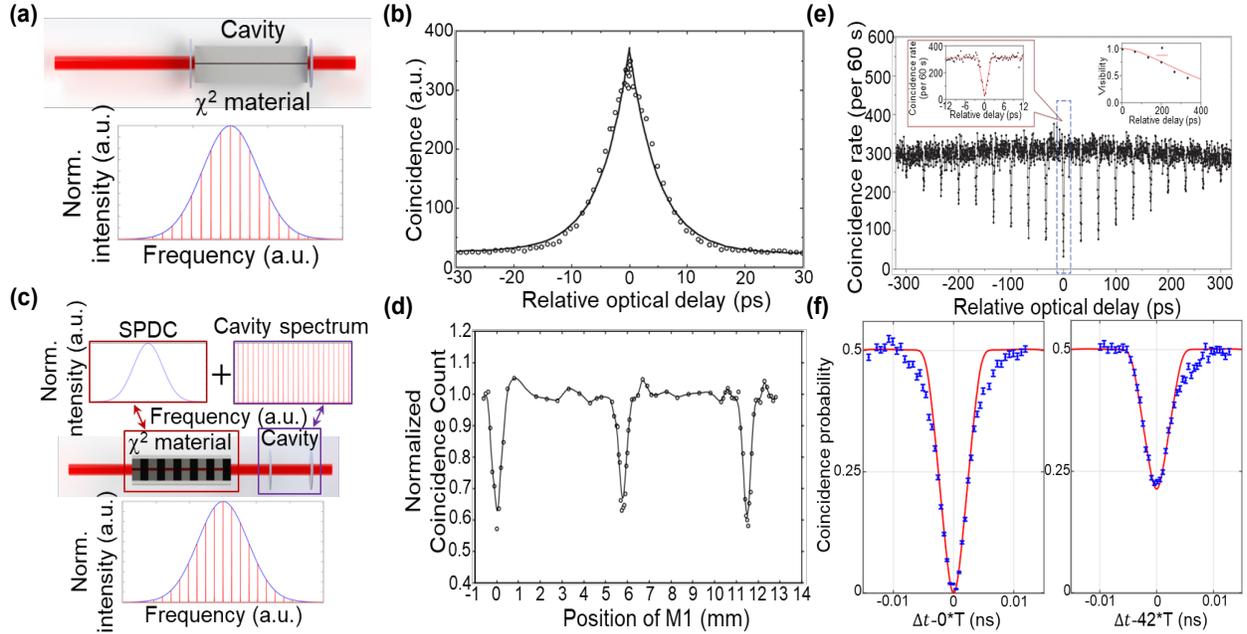

**FIG. 2 Cavity mode-locked mechanism and recurrences of second-order signatures in quantum frequency combs.** (a) The schematic for the generation of mode-locked two-photon states in a Fabry-Perot type cavity-enhanced SPDC. (b) Temporal correlation signatures of single-frequency-mode two-photon states in a cavity-enhanced SPDC. (c) Principles of mode-locked QFC generation using post-filtered SPDC entangled photon pairs. Example SPDC and cavity spectrum, and the output of post-filtered QFC spectrum. The dark red arrow and box are the SPDC spectrum, while the purple arrow and box indicate the spectrum of the filtering cavity. This approach leads to a discrete energy-time correlated QFC. The generated photon pair is entangled across the frequency-bins in the SPDC bandwidth. (d) The first HOM revival measurements of the mode-locked two-photon states. (e) The observed high-visibility 19 HOM revival interferences without post-selection in a mode-locked doubly-filtered QFC. (f) Example HOM revival interferences from a cavity-enhanced SPDC source for 0 and 42 round-trips. In both (e) and (f), the period of HOM revival measurements is half of the cavity round-trip time of the QFCs, resulting from the mode-locking nature of the process. The graphs in panel (b) are reprinted with permission from [75], APS. Panel (d) reprinted with permission from [76], APS. Panel (e) adapted with permission from [79], Springer Nature Limited. Panel (f) adapted with permission from [110], APS.

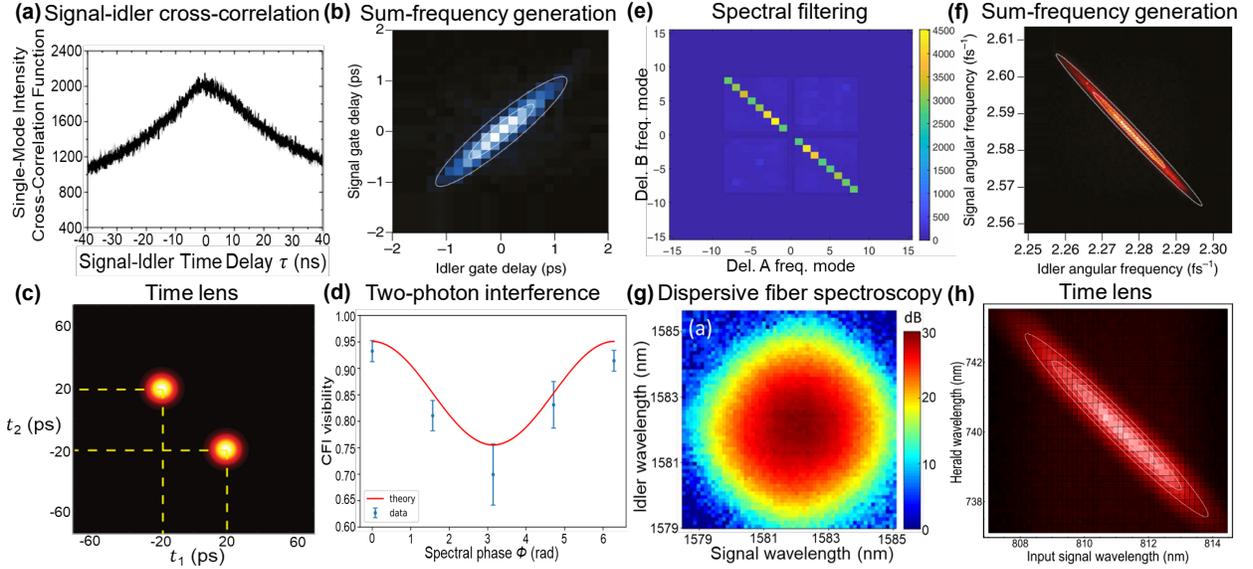

**FIG. 3 Probing joint temporal and spectral characteristics in frequency-correlated SPDC entangled photons.** (a) Temporal second-order cross-correlation function between signal and idler photons in a cavity-enhanced SPDC source. (b) Direct observation of the SPDC JTI with strong positive correlations by using sum-frequency generation (SFG) and ultrafast optical temporal gating pulses. (c) Measured JTI of SPDC photons with the time lens, showing bunched behavior of biphotons. (d) Two-photon conjugate Franson interference visibility versus a function of applied spectral phase, which can be used to probe the JTI between SPDC photons with the same JSI. (e) A measurement of the JSI for an eight-dimensional QFC. (f) Direct measurement of JSI of energy-time entangled SPDC photons using SFG that operate on ultrafast timescales. (g) Measured JSI profile of SPDC biphotons using dispersive fiber spectroscopy. (h) Measured JSI of SPDC photons via time-lens, with strong frequency anticorrelations. The graph in panel (a) is reprinted with permission from [110], AIP. Panel (b) adapted with permission from [114], APS. Panel (c) reprinted from [115], APS. Panel (d) adapted with permission from [122], APS. Panel (e) reprinted from [94], AAAS. Panel (f) reprinted from [114], APS. Panel (g) adapted with permission from [133], OSA. Panel (h) reprinted from [138], APS.

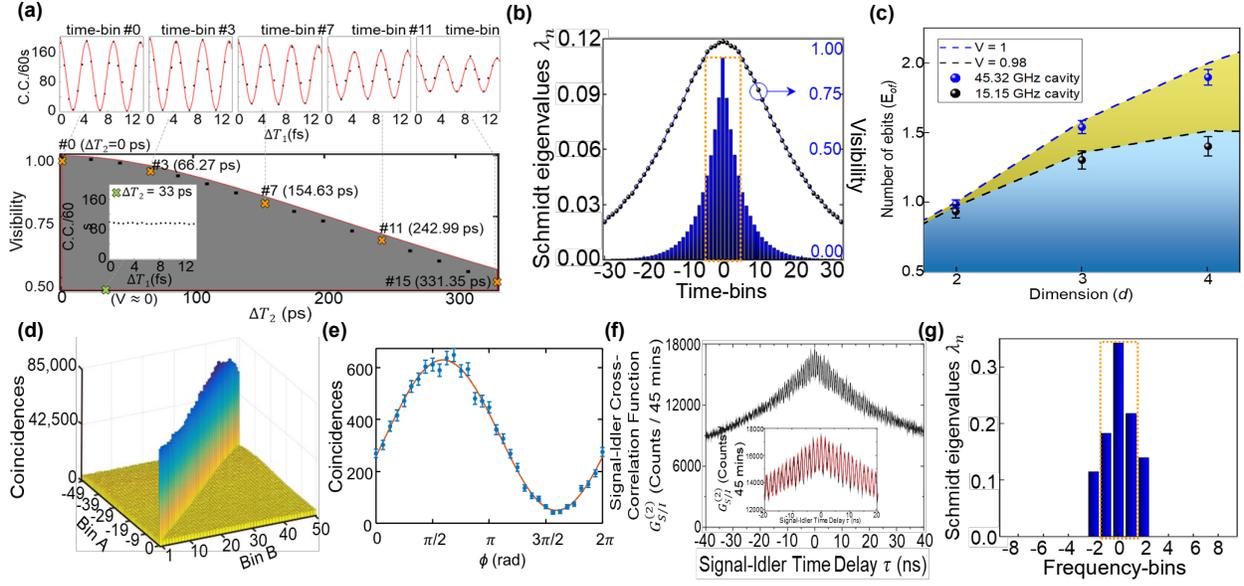

**FIG. 4 High-dimensional time-frequency entanglement in mode-locked quantum frequency combs.** (a) Franson revival interferences using a mode-locked doubly-filtered QFC, demonstrating high-dimensional energy-time entanglement covering 16 time-bins. (b) The Schmidt mode eigenvalues versus different time-bins from HOM interferometry and the corresponding visibilities of the HOM-interference recurrences. (c) Lower bounds for the entanglement of formation (ebits) versus dimension $d$, in reconstructing the density matrix from Franson recurrence interference measurements. (d) Measured frequency-correlated JSI of a mode-locked doubly-filtered QFC, with the measured Schmidt number $K$ about 50. (e) Frequency-domain high visibility interference as a function of spectral phase in a mode-locked QFC. (f) Time-domain second-order cross-correlation measurements using a cavity-enhanced SPDC source. (g) The Schmidt mode eigenvalues for measured QFC states from a 45 GHz FSR cavity. The measurements alone in (d) do not verify high-dimensional frequency-bin entanglement in QFCs, however, by incorporating measurements of frequency-bin mixing and temporal mode-locked oscillations in (e) and (f), they together certify the high-dimensional frequency-bin entanglement in QFC. The graphs in panels (a-c) and (g) are reprinted with permission from [99], Springer Nature Limited. Panel (d) reprinted from [85], OSA. Panel (e) reprinted from [92], Springer Nature Limited. Panel (f) reprinted from [173], APS.

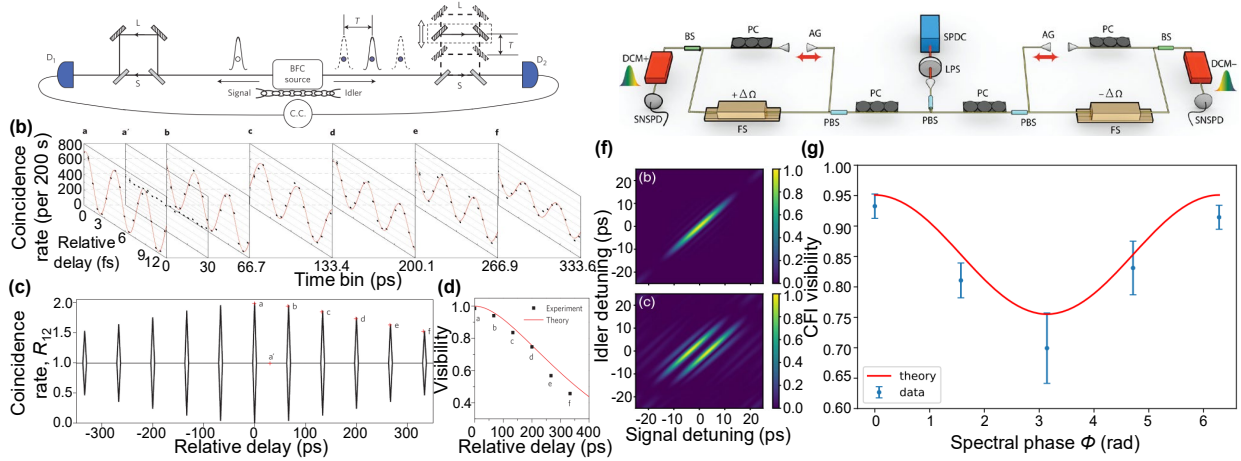

**FIG. 5 Franson and conjugate Franson interferometry for high-dimensional time-frequency entanglement verification.** (a) Experimental scheme of Franson interferometer to verify high-dimensional energy-time entanglement in a QFC state. (b) The first Franson revival interferences measured in a mode-locked QFC, demonstrating high-dimensional energy-time entanglement covering six resonances where a-f indicate discrete temporal positions. (c) Theoretical fringe envelope of Franson interference revivals for a QFC, with superimposed experimental data. (d) Comparison of experimental and theoretical visibility of high-dimensional Franson interference fringes as a function of time delay, with central Franson visibility of 97.8% after accidental subtractions. (e) Experimental setup of the conjugate Franson interferometer, comprising two MZIs with equal-length arms and an optical frequency shifter placed in one arm of each MZI. (f) JTI of a biphoton state with different spectral phases $0/2\pi$ (upper) and $\pi$ (lower). (g) Conjugate Franson fringe visibility as a function of spectral phase. The current experimental conjugate Franson interference visibility is $96 \pm 1\%$ without background subtraction. The conjugate Franson interference visibility degrades when spectral phase variation is introduced. The graphs in panel (a-d) were reprinted with permission from [79], Springer Nature Limited. Panel (e-g) adapted with permission from [122], APS.

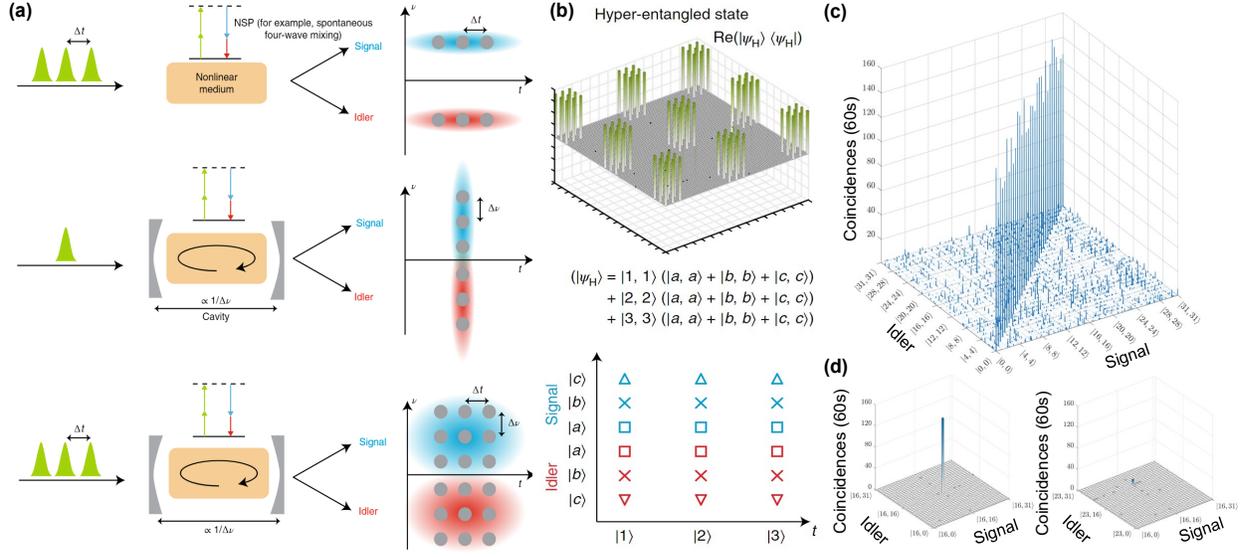

**FIG. 6 Hyperentanglement in mode-locked quantum frequency combs.** (a) Time-frequency hyperentanglement generation scheme. An optical pulse train excites a nonlinear medium to generate a time-bin entangled two-photon state. A single pulse excites a nonlinear medium in a cavity to generate photon-pair over a broad bandwidth, i.e., a frequency-bin entangled state. An optical pulse train excites a nonlinear medium in a cavity to generate a simultaneously time- and frequency-bin entangled state, i.e., a time-frequency hyperentangled state. (b) An example two-photon time-frequency hyperentangled state comprising three temporal modes and three frequency modes per signal and idler photon. (c) The demonstration of a four-party 32-dimensional GHZ state using a high-dimensional gate operation for hyper-entangled QFC, encoding up to 20 qubits. (d) Zoomed-in submatrices of matched and unmatched signal and idler frequency-bins in (c). The graphs in panels (a) and (b) were adapted with permission from [86], Springer Nature Limited. Panel (c) and (d) adapted with permission from [88], Springer Nature Limited.

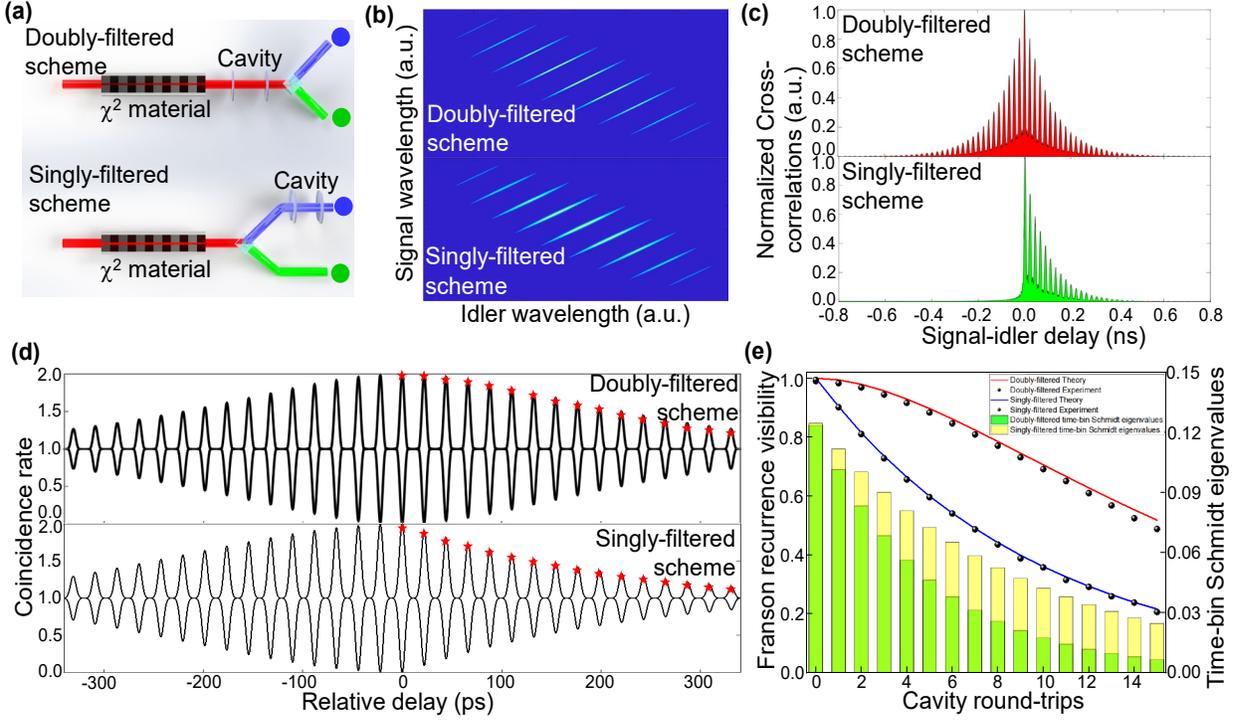

**FIG. 7 Mode-locked doubly- and singly-filtered quantum frequency combs.** (a) Schemes for mode-locked doubly- and singly-filtered QFC. (b) Example JSI of doubly- and singly-filtered QFC. While the FSR of comb structures remains consistent across doubly-filtered and singly-filtered schemes in the frequency domain, their JSI exhibits two notable distinctions. First, in doubly-filtered QFC, the JSI has a more rapid decay as the frequency offset from the comb center increases, contrasting with the singly-filtered scheme. Second, the singly-filtered QFC displays a higher unnormalized power spectral intensity compared to the doubly-filtered configuration. (c) Modeled signal-idler cross-correlation $g^{(2)}$ function (cross-section of JTI) of doubly- and singly-filtered QFC. Here the signal-idler cross-correlation $g^{(2)}$ functions exhibit double- and single-sided temporal oscillating structures with a detection timing jitter $\delta T$ of 5 ps. (d) The experimental and theoretical visibilities of the Franson revival interference fringes in mode-locked doubly- and singly-filtered QFC. The coincidence counts in Franson interference experiments are superimposed for 16 positive time-bins. (e) The Franson recurrence visibility and corresponding time-bin Schmidt eigenvalues for mode-locked doubly- (red line and yellow bar) and singly-filtered (blue line and green bar) QFCs. The graphs and data in panels (d) and (e) are reprinted with permission from [99], and [149], Springer Nature Limited.

# TABLE 1. Comparison of recent quantum frequency combs.

| Process | Platform | Configuration | $\lambda$ (nm) | Encoding | Photons | Frequency modes | Temporal modes | Schmidt number | Ebits | Focus | Reference |
|---|---|---|---|---|---|---|---|---|---|---|---|
| SPDC | PPKTP | Doubly-filtered | 1316 + 1316 | Energy-time-polarization | 2 | 10 | 19 | / | / | High-dimensional hyperentanglement QFC | [79] |
| SPDC | PPKTP | DR | 795 + 795 | Time-bin | 2 | ~800 | 84 | / | / | Hectometer HOM revivals in a QFC | [109] |
| SFWM | Glass | DR | ~1550 | Time-frequency | 2-4 | 10 | 10 | 10.45 ± 0.53 | / | On-chip generation and coherent control of high-dimensional QFC | [82] |
| SFWM | $Si_3N_4$ | DR | ~1550 | Time-frequency | 2 | 40 | / | 20 | / | 50 GHz spacing on-chip QFC | [84] |
| SPDC | PPLN | Doubly-filtered | ~1550 | Time-frequency | 2 | 50 | / | / | / | Frequency-bin HOM interference in a QFC | [85] |
| SPDC | PPLN | Doubly-filtered | ~1550 | Time-frequency | 2 | 50 | / | / | / | Frequency-bin CNOT gate using a QFC | [87] |
| SPDC + SFWM | PPLN + $Si_3N_4$ | Doubly-filtered + DR | ~1550 | Time-frequency | 2 | 256 | 256 | / | 1.19 ± 0.12 | Bipartite GHZ state with a Hilbert space of 20 qubits in a QFC | [88] |
| SPDC | PPLN | SR | 1520-1600 | Time-frequency | 2 | 1000 | ~30 | / | / | A broadband multimode QFC | [90] |
| SPDC | PPLN | Doubly-filtered | ~1550 | Time-frequency | 2 | 50 | / | / | / | High-dimensional quantum walk with a QFC | [94] |
| SPDC | AlGaAs | Doubly-filtered | ~1530 | Time-frequency | 2 | 500 | / | / | / | Integrated semiconductor QFC source | [97] |
| SPDC | PPKTP | Doubly-filtered | 1316 + 1316 | Time-frequency-polarization | 2 | 19 | 61 | 18.30 | 1.89 ± 0.03 | A 648 Hilbert space dimensional QFC | [99] |
| SPDC | PPLN | SR | 1520-1600 | Time-frequency | 2 | 1400 | ~30 | 10 | / | Massive-mode polarization entangled QFC | [100] |
| SPDC + SFWM | PPLN + $Si_3N_4$ | Doubly-filtered + DR | ~1550 | Time-frequency | 2 | 51 | / | / | 2.50 ± 0.08 | Frequency-bin quantum state tomography using QFCs | [92] |
| SPDC | PPKTP | Singly-filtered | 1316 + 1316 | Time-frequency | 2 | 5 | 16 | 13.11 | / | High-dimensional entanglement distribution and QKD with singly-filtered QFC | [149] |

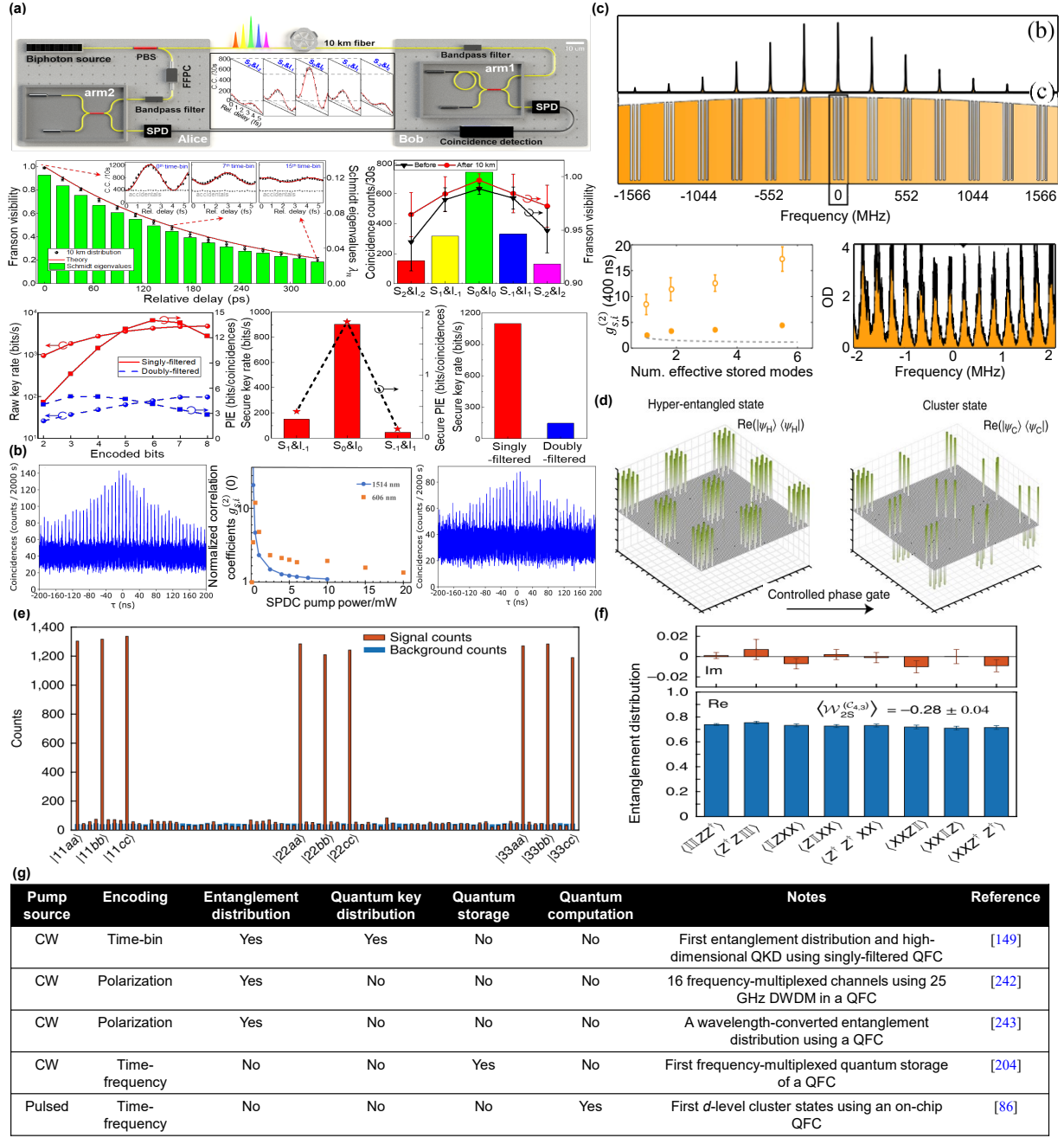

**FIG. 8 Quantum communications and computation processing with mode-locked quantum frequency combs.** (a) An efficient entanglement distribution using time- and frequency-bins of a mode-locked singly-filtered QFC after a 10-km fiber link. Proof-of-principle high-dimensional QKD using mode-locked singly- and doubly-filtered QFC has also been demonstrated. (b) The temporal multimode nature of a wavelength-converted DR-OPO with 10-km fiber. (c) Matched atomic frequency comb and DR-OPO spectrums, and the measured second-order cross-correlation function for stored frequency-multiplexed QFC. (d) Generation of $d$-level cluster states from time-frequency hyperentangled state with a controlled phase gate. (e) Measured photon projections on the 81 diagonal elements of the cluster state density matrix. (f) Real (blue bars) and imaginary (red

bars) parts of the measured expectation values for the individual terms of the cluster state witness operator. (g) A summary table for several recent studies using QFC towards quantum communications and computation. The graph in panel (a) is reprinted with permission from [149], Springer Nature Limited. Panel (b) adapted with permission from [243], Springer Nature Limited. Panel (c) reprinted with permission from [204], APS. Panel (d-f) adapted from [86], Springer Nature Limited.